\documentclass[a4paper,11pt]{article}

\usepackage{jcappub} 

\usepackage[T1]{fontenc} 

\newcommand{\bd}[1]{\mbox{\boldmath$#1$}}

\title{The optical geometry definition of the total deflection angle 
of a light ray in curved spacetime}

\author[a]{Hideyoshi Arakida\note{Corresponding author.}}

\affiliation[a]{College of Engineering, Nihon University,\\
Koriyama, Fukushima 963-8642 JAPAN}

\emailAdd{arakida.hideyoshi@nihon-u.ac.jp}

\abstract{
Assuming a static and spherically symmetric spacetime,
we propose a novel concept of the total deflection angle 
of a light ray in terms of the optical geometry which is the 
Riemannian geometry experienced by the light ray. 
The total deflection angle is defined by the difference between 
the sum of internal angles of two triangles; one of the triangles lies 
on curved spacetime distorted by a gravitating body and the other 
on its background. The triangle required to define the total 
deflection angle can be realized by setting three laser-beam 
baselines as in planned space missions such as LATOR, ASTROD-GW, 
and LISA. Accordingly, the new total deflection angle is, in principle, 
measurable by gauging the internal angles of the triangles. 
The new definition of the total deflection angle can provide 
a geometrically and intuitively clear interpretation. 
Two formulas are proposed to calculate the total deflection angle 
on the basis of the Gauss--Bonnet theorem.
It is shown that in the case of the Schwarzschild spacetime,
the expression for the total deflection angle $\alpha_{\rm Sch}$
reduces to Epstein--Shapiro's formula when the source of a light ray 
and the observer are located in an asymptotically flat region. 
Additionally, in the case of the Schwarzschild--de Sitter spacetime, 
the expression for the total deflection angle $\alpha_{\rm SdS}$ 
comprises the Schwarzschild-like parts and coupling terms of 
the central mass $m$ and the cosmological constant $\Lambda$ in 
the form of ${\cal O}(\Lambda m)$ instead of ${\cal O}(\Lambda/m)$.
Furthermore, $\alpha_{\rm SdS}$ does not include the terms characterized 
only by the cosmological constant $\Lambda$.
}

\usepackage{amsmath}	
\begin{document}
\maketitle
\flushbottom

\section{Introduction\label{sec:intro}}
Generally, the cosmological constant $\Lambda$, or dark energy, 
is considered a promising candidate that can explain the accelerating 
expansion of the universe \cite{riess1998,schmidt1998,perlmutter1999}. 
Although intensive studies have been performed for solving the mystery 
surrounding the cosmological constant/dark energy  
from both the theoretical and observational viewpoints, 
definitive direct evidence has not been obtained.

Because the structure of spacetime (the form of the metric $g_{\mu\nu}$) 
depends on the existence of the cosmological constant $\Lambda$, 
the evidence of the existence of the cosmological constant may 
be detected using the classical tests of general relativity, 
such as the observation of light deflection and the perihelion/periastron 
advance. Especially, light deflection is the basis of gravitational lensing 
which is a powerful tool in astrophysics and cosmology; for more details, 
see \cite{schneider_etal1999,schneider_etal2006} and the references therein. 
Therefore, gravitational lensing may prove the existence of the cosmological 
constant.

Thus far, the influence of the cosmological constant on light 
deflection has been studied in various ways, including 
whether the cosmological constant contributes to light deflection. 
Historically, Islam \cite{islam1983} first mentioned that light trajectory 
is independent of the cosmological constant $\Lambda$ because the second-order 
differential equation of the light ray does not include $\Lambda$. Therefore, 
it was considered for a long time that $\Lambda$ did not contribute 
to light deflection. However, in 2007, Rindler and Ishak 
\cite{rindler2007} indicated that $\Lambda$ affects the bending 
of a light ray, by using the invariant cosine formula under the 
Schwarzschild--de Sitter/Kottler solution. Starting with this paper 
\cite{rindler2007}, many authors intensively discussed its appearance  
in diverse ways; see \cite{ishak2010} for a review article, 
and also see 
\cite{lake2002,park2008,kp2008,sph2010,bhadra2010,miraghaei2010,biressa2011,ak2012,hammad2013,lebedevlake2013,batic_etal2015,arakida2016} 
and the references therein.

Despite intensive research in the past, a definitive conclusion 
has not been drawn, mainly because of the following reasons:
\begin{itemize}
 \item Unlike the Schwarzschild spacetime, the spacetime does not become 
       asymptotically flat because of the cosmological 
       constant $\Lambda$. Accordingly, we cannot apply the 
       standard procedure, which is described in many textbooks and 
       literature, for calculating the total deflection angle.
 \item \cite{rindler2007} indicated that to calculate the angle 
       in curved spacetime, one must focus on not only the equation 
       of light trajectory but also the metric (ruler) of spacetime. 
       However, in many studies, only the equation of light trajectory 
       was considered when discussing the total deflection angle.
 \item It is still not clear what is the total deflection angle and 
       how it should be defined in curved spacetime.
\end{itemize}
Especially, the third reason above seems to be an essential problem that
makes it difficult to clarify the contribution of 
the cosmological constant to total deflection angle. 
To overcome this difficulty, some authors applied the 
the Gauss--Bonnet theorem 
\cite{gibbons_werner2008a,ishihara_etal2016,ishihara_etal2017,arakida2018,takizawa_etal2020}.
Although the Gauss--Bonnet theorem might help in 
solving the problem of the total deflection angle, it has still not been 
resolved and thus further consideration is necessary. 

This study aims to provide a renewed method to solve 
the problem of total deflection angle and reveal the influence 
of the cosmological constant on light deflection. 
To this end, by assuming a static and spherically symmetric spacetime,
we propose a new concept of the total deflection angle of a light ray
in terms of the optical geometry which is regarded as the Riemannian 
geometry experienced by the light ray; 
the concept is realized by considering the difference between 
the sum of internal angles of two triangles; one exists in curved spacetime 
distorted by a gravitating body and the other in its background.  
This concept of the total deflection angle is inspired by space missions 
including LATOR \cite{lator2009}, ASTROD-GW \cite{astrodgw}, and 
LISA \cite{lisa}, which set three laser-beams baselines in the space; 
accordingly, the new total deflection angle could be measured, 
in principle, by gauging the internal angle of each triangle. 
The new total deflection angle is geometrically and intuitively clear 
to interpret. To calculate the total deflection angle, we develop 
two formulas using the Gauss--Bonnet theorem.

This paper is outlined as follows. Section \ref{sec:optical}
introduces the optical metric that is considered the 
Riemannian geometry of light rays. Section \ref{sec:GBtheorem}
summarizes the Gauss--Bonnet theorem.
Section \ref{sec:newangle} proposes a new concept and definition 
of the total deflection angle and presents two formulas to calculate
the total deflection angle.
In Sections \ref{sec:Schwarzschild} and \ref{sec:Schwarzschild-deSitter},
we apply the two formulas to compute the total deflection angle
in the Schwarzschild and Schwarzschild--de Sitter spacetimes,
respectively. Finally, conclusions are drawn in Section \ref{sec:conclusion}.
\section{Optical Metric\label{sec:optical}}
We assume the following static and spherically symmetric spacetime:
\footnote{
It is possible to start from more generic spherical metric form as
$ds^2 = -A(r)dt^2 + B(r)dr^2 + C(r)(d\theta^2 + \sin^2 \theta d\phi^2)$,
and to construct the optical metric, see e.g., 
\cite{ishihara_etal2016,ishihara_etal2017,takizawa_etal2020}.
However, in this paper, we adopt a metric of type $g_{tt} = -1/g_{rr}$ 
because we are interested in the propagation of light in Schwarzschild 
and Schwarzschild--de Sitter spacetimes.}
\begin{align}
 ds^2 &= g_{\mu\nu}dx^{\mu}dx^{\nu}
  \nonumber\\
  &=  -f(r)dt^2 + \frac{1}{f(r)}dr^2
  + r^2(d\theta^2 + \sin^2\theta d\phi^2),
  \label{eq:metric1}
\end{align}
where $g_{\mu\nu}$ denotes the metric tensor of spacetime whose signature 
is denoted by $\eta_{\mu\nu} = {\rm diag}(-: +: +: +)$; 
$f(r)$ denotes a function of the radial coordinate $r$, 
Greek indices e.g., $\mu, \nu$, range from 0 to 3, and we choose 
the geometrical unit $c = G = 1$ throughout this paper. 
Because of the spherical symmetry, we consider the equatorial plane 
$\theta = \pi/2, d\theta = 0$ as the orbital plane of the light rays. 
One has 
\begin{align}
 ds^2 = -f(r)dt^2 + \frac{1}{f(r)}dr^2 + r^2 d\phi^2.
  \label{eq:metric2}
\end{align}
Two constants of motion, energy $E$ and angular momentum $L$, 
are expressed as
\begin{align}
 E = f(r)\frac{dt}{d\lambda},\quad
  L = r^2\frac{d\phi}{d\lambda},
  \label{eq:constant1}
\end{align}
where $\lambda$ denotes an affine parameter. Furthermore, 
the impact parameter $b$ is introduced as
\begin{align}
 b \equiv \frac{L}{E}.
  \label{eq:impact}
\end{align}

Using the null condition $ds^2 = 0$, we introduce the optical 
metric $\bar{g}_{ij}$, which is considered the Riemannian geometry
experienced by light rays. One has 
\begin{align}
 dt^2 &\equiv \bar{g}_{ij} dx^idx^j
  = \left(\frac{g_{ij}}{g_{00}}\right) dx^idx^j\nonumber\\
 &= \bar{g}_{rr}dr^2 + \bar{g}_{\phi\phi}d\phi^2
  \nonumber\\
 &= \frac{1}{[f(r)]^2}dr^2 + \frac{r^2}{f(r)}d\phi^2,
  \label{eq:optical1}
\end{align}
where Latin indices, e.g., $i, j$, assume the values of $i, j = 1, 2$, 
which correspond to $1 = r$ and $2 = \phi$, respectively. 
Hereafter, in accordance with \cite{abramowicz1988}, 
we refer to the geometry defined by the optical metric 
$\bar{g}_{ij}$ as the optical (reference) geometry ${\cal M}^{\rm opt}$. 
In the optical geometry description of Einstein's relativity, 
light trajectories in the 3-dimensional space corresponding to a static 
spacetime are geodesic lines. This allows one to give a precise and practical 
definition of total deflection angle of the light ray.

Notably, on the optical reference geometry ${\cal M}^{\rm opt}$, 
it is observed from Eq. (\ref{eq:optical1}) that the time coordinate 
$t$ plays the role of an arc length parameter because
\begin{align}
 \int_{t_1}^{t_2} dt =
  \int_{t_1}^{t_2}
  \sqrt{\bar{g}_{rr}(k^r)^2 + \bar{g}_{\phi\phi}(k^{\phi})^2}dt
  = t_2 - t_1,
  \label{eq:optical3}
\end{align}
where $k^i = dx^i/dt$ denotes the unit tangent vector along the path of 
the light ray on ${\cal M}^{\rm opt}$, and it satisfies  
$1 = \bar{g}_{ij}k^i k^j$ from Eq. (\ref{eq:optical1}).
The property given in Eq. (\ref{eq:optical3}) is appropriate for 
application to the Gauss--Bonnet theorem later.

Let us summarize some important properties of the optical metric 
$\bar{g}_{\mu\nu}$. If considering the slice of constant time $t$ of 
the spacetime Eq. (\ref{eq:metric2}), the spatial part of the metric $g_{ij}$ 
is described as
\begin{align}
 d\ell^2 &\equiv g_{ij}dx^idx^j
  \nonumber\\
 &=
  g_{rr}dr^2 + g_{\phi\phi}d\phi^2
  \nonumber\\
 &=
  \frac{dr^2}{f(r)} + r^2d\phi^2.
  \label{eq:metric3}
\end{align}
Two metrics, Eqs. (\ref{eq:optical1}) and (\ref{eq:metric3}), 
are connected by the conformal transformation (conformal mapping) as
\begin{align}
 \bar{g}_{ij} = \omega^2 (\bd{x}) g_{ij},\quad
  \omega^2 (\bd{x}) = \frac{1}{f(r)},
  \label{eq:metric4}
\end{align}
or more generally,
\begin{align}
  \bar{g}_{\mu\nu} = \omega^2 (\bd{x})g_{\mu\nu},
\end{align}
where $\omega^2(\bd{x})$ denotes the conformal factor. Because
the conformal transformation preserves the angle of the point at 
which the two curves intersect, the angles remain the same 
in both $\bar{g}_{ij}$ and $g_{ij}$.
However, the conformal transformation rescales the coordinate value.
Moreover, the null geodesic does not change its form 
upon performing the conformal transformation because of the null condition
$ds^2 = 0$; see Appendix G in \cite{carroll2004}.
\section{Gauss--Bonnet Theorem\label{sec:GBtheorem}}
\begin{figure}[htbp]
\begin{center}
 \includegraphics[scale=0.2,clip]{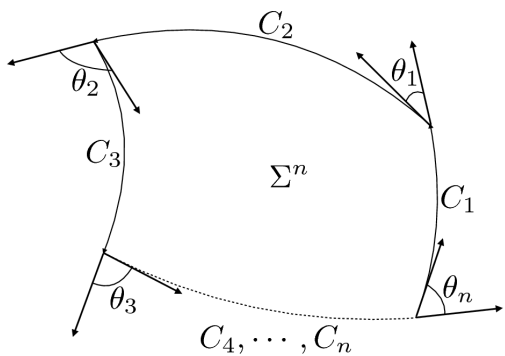}
 \caption{ Gauss--Bonnet theorem.
 A polygon $\Sigma^n$ is bounded by curves $C_1, C_2, \cdots, C_n$,
 and the external angles of polygon $\Sigma^n$ are denoted by
 $\theta_p ~ (p = 1, 2, \cdots, n)$.\label{fig:arakida-fig1}}
\end{center}
\end{figure}

In the optical reference geometry ${\cal M}^{\rm opt}$ given by 
the metric $\bar{g}_{ij}$, (see Eq. (\ref{eq:optical1})), 
we consider an $n$-vertex polygon $\Sigma^n$, which is orientable 
and bounded by $n$ smooth and piecewise regular curves 
$C_p ~ (p = 1, 2, \cdots, n)$ (see Figure \ref{fig:arakida-fig1}).
The (local) Gauss--Bonnet theorem is expressed as described 
in, e.g., on p. 139 
in \cite{klingenberg1978}, p. 170 in \cite{kreyszig1991}, and 
p. 272 in \cite{carmo2016}, as follows:
\begin{align}
 \iint_{\Sigma^n} Kd\sigma
  +
 \sum_{p = 1}^n \int_{C_p} \kappa_g dt
  + \sum_{p = 1}^{n}\theta_p = 2\pi,
  \label{eq:GB1}
\end{align} 
where an arc length parameter is denoted by $t$ instead of $s$
(see Eq. (\ref{eq:optical3})), and an arc length parameter $t$ moves 
along the curve $C_p$ in such a manner that a polygon $\Sigma^n$ 
stays on the left side; $\theta_p$ denotes the external angle at the 
$p$-th vertex, and $\theta_p$ is determined as the sense leaving 
the internal angle on the left. 
$K$ denotes the Gaussian curvature as defined in, e.g., p. 147
of \cite{kreyszig1991}
\begin{align}
 K = -\frac{1}{\sqrt{\bar{g}_{rr} \bar{g}_{\phi\phi}}}
  \left[
   \frac{
   \partial}{\partial r}
   \left(\frac{1}{\sqrt{\bar{g}_{rr}}}
    \frac{\partial\sqrt{\bar{g}_{\phi\phi}}}{\partial r}\right)
   +
   \frac{\partial}{\partial \phi}
   \left(\frac{1}{\sqrt{\bar{g}_{\phi\phi}}}
    \frac{\partial\sqrt{\bar{g}_{rr}}}{\partial \phi}\right)
	 \right],
  \label{eq:GB2}
\end{align}
which represents the manner in which the spacetime is curved, and
$d\sigma = \sqrt{|\det(\bar{g}_{ij})|}dx^1 dx^2 
= \sqrt{|\det(\bar{g}_{ij})|}drd\phi$
denotes an areal element. 
The term $\kappa_g$ denotes the geodesic curvature along 
curve $C_p$, e.g., on p. 256 in \cite{carmo2016} 
\begin{align}
 \kappa_g = \frac{1}{2\sqrt{\bar{g}_{rr} \bar{g}_{\phi\phi}}}
  \left(
   \frac{\partial \bar{g}_{\phi\phi}}{\partial r}
   \frac{d\phi}{dt}
   -
   \frac{\partial \bar{g}_{rr}}{\partial \phi}
   \frac{dr}{dt}
	 \right)
  + \frac{d\Phi}{dt},
  \label{eq:GB3}
\end{align}
where $\Phi$ denotes the angle between the radial unit vector $e_r^i$
along radial geodesics and the tangent vector $k^i = dx^i/dt$ 
of curve $C_p$. The term $\kappa_g$ characterizes the extent 
to which curve $C_p$ deviates from the geodesic. 
Accordingly, if curve $C_p$ is the geodesic, $\kappa_g = 0$. 
\section{New Concept of the Total Deflection Angle
\label{sec:newangle}}
We propose a new concept of the total deflection angle $\alpha$. 
Intuitively, the total deflection angle of 
a light ray is the change in the direction of light ray
in the presence and absence of a gravitating body. 
However, the metric of spacetime is different in the presence and 
absence of gravitating body, thus each null geodesic essentially 
exists in a distinct spacetime determined by a different metric 
(ruler); for instance, in the 
case of the Schwarzschild spacetime, one null geodesic lies on curved 
spacetime (curved ruler), while the another exists in 
flat spacetime (flat ruler). Therefore, it becomes difficult to 
compare two null geodesics with each other in the same spacetime 
(same ruler). 
As an exception, the total deflection angle can be obtained only 
when both observer $R$ and light source $S$ are placed in 
asymptotically flat regions, owing to the Euclidean parallel postulate, 
which enables us to determine the angle at a distant point from 
the angle of any point $P$, such as the corresponding angle and
the alternate angle.
Notably, the total deflection angle in the Schwarzschild spacetime 
can be obtained as the twice of angle $\psi_P$ at $P$, i.e., 
$\alpha = 2\psi_P$,
where $P$ denotes the light source $S$ or observer $R$. 
However, in a curved spacetime or region, the parallel postulate 
does not hold, and thus $\alpha \ne 2\psi_P$.

However, even in curved spacetime, the internal angles of polygon 
$\Sigma^n$ can be measured by using the equipment mounted on the
spacecrafts located at each vertex. Accordingly, the sum of 
the internal angles of polygon $\Sigma^n$ can be calculated. 
Because the sum of the internal angles of the polygon depends 
on the curvature of the spacetime, 
one might consider the difference between the sum of 
internal angles of two polygons that placed in distinct spacetimes. 
Therefore, we define the renewed total deflection angle $\alpha$ as 
the difference between the sums of the internal angles of two polygons.

\begin{figure}[htbp]
\begin{center}
 \includegraphics[scale=0.4,clip]{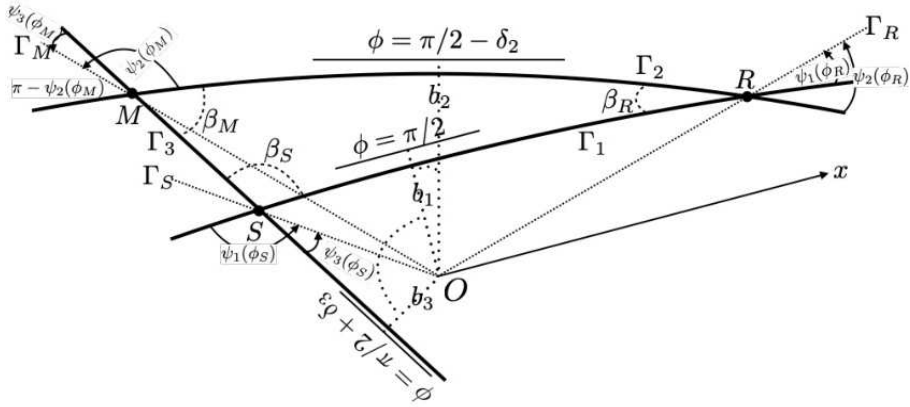}
 \caption{Configuration of triangle $\Sigma^3$.
 A triangle $\Sigma^3$ is bounded by three null geodesics,
 $\Gamma_1$, $\Gamma_2$, and $\Gamma3$, whose impact parameters 
 are denoted by $b_1$, $b_2$, and $b_3$, respectively.
 $R$, $M$, and $S$ denote the triangle vertexes where
 the satellites or ISS/ground stations are located.
 For simplicity, the three null geodesics are arranged 
 such that the point of the closest approach corresponds
 to $\phi = \pi/2$, $\phi = \pi/2 - \delta_2$, and $\phi = \pi/2 + \delta_3$,
 respectively. $\Gamma_R$, $\Gamma_M$, and $\Gamma_S$ 
 correspond to the coordinate lines
 $\phi = \phi_R$, $\phi = \phi_M$, and $\phi = \phi_S$, respectively 
 passing through the points $R$, $M$, and $S$, respectively.
 \label{fig:arakida-fig2}}
\end{center}
\end{figure}

To feasibly realize the above-mentioned concept of total deflection angle, 
we construct triangle $\Sigma^3$ on the optical reference geometry 
${\cal M}^{\rm opt}$; the triangle is bounded by three null geodesics
$\Gamma_1$, $\Gamma_2$, and $\Gamma_3$. Practically, these
three null geodesics can be substantiated by three laser-beam 
baselines that connect three spacecrafts, or two spacecrafts and 
International Space Station/ground station on the Earth,
as in planned missions, e.g., LATOR, ASTROD-GW, and LISA. 
For more details, see Figure \ref{fig:arakida-fig2}, wherein $R$, $M$, 
and $S$ denote the triangle vertexes where the satellites or ISS/ground 
stations are located. We denote the impact parameters of three 
null geodesics $\Gamma_1$, $\Gamma_2$, and $\Gamma_3$ by $b_1$, $b_2$, 
and $b_3$, respectively. For simplicity, we arrange 
triangle $\Sigma^3$ such that the angular coordinates 
$\phi$ at the closest approach of $\Gamma_1$, $\Gamma_2$, and $\Gamma_3$ 
correspond to $\phi = \pi/2$, $\phi = \pi/2 - \delta_2$, and 
$\phi = \pi/2 + \delta_3$, respectively. $\Gamma_R$, $\Gamma_M$, and
$\Gamma_S$ 
correspond to the coordinate lines
$\phi = \phi_R$, $\phi = \phi_M$, and $\phi = \phi_S$, respectively 
passing through the points $R$, $M$, and $S$, respectively.
$\Gamma_R$, $\Gamma_M$, and $\Gamma_S$ are related to the unit radial 
vector in the equatorial plane, noting that we are working in the 
optical reference geometry ${\cal M}^{\rm opt}$,
\begin{align}
 e^i_r = \left(\frac{1}{\sqrt{\bar{g}_{rr}}}, 0\right).
\end{align}
We note that for the observer at the points $R$, $M$, and $S$, 
$\Gamma_R$, $\Gamma_M$, and $\Gamma_S$ can be regarded as the radial null 
geodesics of the light rays coming from the central object $O$
\footnote{
It may be noteworthy that in terms of the null geodesics, 
$\Gamma_1$, $\Gamma_2$, $\Gamma_3$, $\Gamma_R$, $\Gamma_M$, and $\Gamma_S$ 
are more precisely spatial geodesics which are the projections of 
null geodesics on the spatial spacetime sections $t = {\rm const}$.
However, due to the nature of conformal transformation, the form of the 
differential equation of the photon trajectory $dr/d\phi$ on the optical 
metric is the same as the form of the equation on the spatial spacetime 
sections $t = {\rm const}$ derived from the null condition $ds^2 = 0$.},
however, from the point of view of the coordinate system, $\Gamma_R$, $\Gamma_M$, 
and $\Gamma_S$ are not radial because the observer cannot be at $r = 0$.

Let us define the renewed total deflection angle in accordance
with the above-mentioned configuration of triangle $\Sigma^3$.
Because $\Gamma_1$, $\Gamma_2$, and $\Gamma_3$ are null geodesics, 
the line integral of geodesic curvature $\kappa_g$ vanishes. One has
\begin{align}
 \int_{C_p} \kappa_g dt = 
 \int_{\Gamma_p} \kappa_g dt = 0, \quad p = 1, 2, 3.
\end{align}
Accordingly, the Gauss--Bonnet theorem, Eq. (\ref{eq:GB1}), reduces to
\begin{align}
 \iint_{\Sigma^3}Kd\sigma + \sum_{p = 1}^{3}\theta_p = 2\pi.
  \label{eq:GBcurved}
\end{align}
Next, we prepare the same relation as that in Eq. (\ref{eq:GBcurved})
for the background spacetime, e.g., Minkowski and de Sitter 
spacetimes, and write the following:
\begin{align}
 \iint_{\Sigma^3}K^{\rm BG}d\sigma^{\rm BG} +
  \sum_{p = 1}^{3}\theta^{\rm BG}_p = 2\pi,
  \label{eq:GBback}
\end{align}
where superscript ${\rm BG}$ denotes the background. Additionally, 
for the same reason as that in the case of Eq. (\ref{eq:GBcurved}), 
the line integral of geodesic curvature $\kappa_g^{\rm BG}$ is also zero. 

Rewriting the sum of the external angles, $\theta_p$, using
the internal angles $\beta_p$, we have
\begin{align}
 \sum_{p = 1}^{3} \theta_p &= 3\pi - \sum_{p = 1}^{3}\beta_p,
  \label{eq:ext_int}\\
 \sum_{p = 1}^{3} \theta_p^{\rm BG} &= 3\pi - \sum_{p = 1}^{3}\beta^{\rm BG}_p,
  \label{eq:ext_intbg}
\end{align}
where we used the relation $\theta_p = \pi - \beta_p$.
Subtracting Eq. (\ref{eq:GBback}) from Eq. (\ref{eq:GBcurved}) and
using Eqs. (\ref{eq:ext_int}) and (\ref{eq:ext_intbg}),
we obtain the following relation:
\begin{align}
 \iint_{\Sigma^3}Kd\sigma
  - \
  \iint_{\Sigma^3}K^{\rm BG}d\sigma^{\rm BG}
  =
  \sum_{p = 1}^{3}
  \left(\beta_p - \beta^{\rm BG}_p\right).
  \label{eq:deftotalangle1}
\end{align}
Using the right-hand side of Eq. (\ref{eq:deftotalangle1}),
let us define the renewed total deflection angle as
\begin{align}
 \alpha \equiv
  \left|
   \sum_{p = 1}^{3}
  \left(\beta_p - \beta^{\rm BG}_p\right)
 \right|.
  \label{eq:deftotalangle2}
\end{align}
Eq. (\ref{eq:deftotalangle2}) represents the difference in the
sum of internal angles $\beta_p$ between two triangles. 
Eq. (\ref{eq:deftotalangle2}) provides an instinctive and clear definition 
of the total deflection angle. The absolute-value symbol in 
Eq. (\ref{eq:deftotalangle2}) indicates 
that we take the total deflection angle to be a positive value.

To obtain the internal angles $\beta_p$, we compute
$\psi_p$, which denote the intersection angles between the
null geodesics $(\Gamma_p ~~ (p = 1, 2, 3))$ and radial
null geodesics $(\Gamma_R$ $\Gamma_M$, and $\Gamma_S)$. 
Angles $\psi_p$ can be calculated using the tangent formula as follows:
\begin{align}
 \tan\psi_p =
  \frac{\sqrt{\bar{g}_{\phi\phi}(r_p)}}
  {\sqrt{\bar{g}_{rr}(r_p)}}
  \frac{d\phi}{dr_p}
  =
  \sqrt{f(r_p)}r_p
  \frac{d\phi}{dr_p},\quad
  p = 1, 2, 3.
  \label{eq:psi1}
\end{align}
Using $\psi_p$, the internal angle can be calculated as
\begin{align}
 \beta_R &= \psi_2(\phi_R) - \psi_1(\phi_R),
  \label{eq:betaR}\\
 \beta_M &= \psi_3(\phi_M) - \psi_2(\phi_M) + \pi,
  \label{eq:betaM}\\
 \beta_S &= \psi_1(\phi_S) - \psi_3(\phi_S),
  \label{eq:betaS}
\end{align}
where the subscripts, $1$, $2$, and $3$, of $\psi$ 
represent the intersection angle between light trajectory,
$\Gamma_1$, $\Gamma_2$, and $\Gamma_3$, and radial null geodesics.
Additionally, $\phi_S$, $\phi_M$, and $\phi_R$ denote the angular
coordinate values at points $S$, $M$, and $R$, respectively.
See Figure \ref{fig:arakida-fig2}. 
Notably, angle $\psi_p$ is determined to be in the counterclockwise direction
from the light trajectory, $\Gamma_1$, $\Gamma_2$, and $\Gamma_3$  
to the radial geodesics, $\Gamma_R$, $\Gamma_M$, and $\Gamma_S$, 
(see Figure \ref{fig:arakida-fig2}).
Notably, Eq. (\ref{eq:psi1}) gives 
angle $\psi_p$ by considering the metric, namely, 
the part $\sqrt{f(r_p)}$, in Eq. (\ref{eq:psi1}). 

Using Eq. (\ref{eq:deftotalangle1}), the renewed total deflection angle
$\alpha$ can be also expressed as the difference in the areal integral 
of the Gaussian curvature $K$ as
\begin{align}
 \alpha \equiv \left|
	   \iint_{\Sigma^3}Kd\sigma
	   - \iint_{\Sigma^3}K^{\rm BG}d\sigma^{\rm BG}
	 \right|.
 \label{eq:deftotalangle4}
\end{align}
We refer to Eq. (\ref{eq:deftotalangle2}) as the {\it angular formula of
the total deflection angle} and Eq. (\ref{eq:deftotalangle4}) 
the {\it integral formula of the total deflection angle}.
\section{Total Deflection Angle in the Schwarzschild Spacetime
\label{sec:Schwarzschild}}
We first examine the total deflection angle in the Schwarzschild 
spacetime as
\begin{align}
 f^{\rm Sch}(r) = 1 - \frac{2m}{r},
\end{align}
where $m$ denotes the mass of the central (lens) object. 
\subsection{Light Trajectory}
The first-order differential equation for null geodesic is given as
\begin{align}
 \left(\frac{dr^{\rm Sch}_p}{d\phi}\right)^2
  = 
  (r^{\rm Sch}_p)^2\left[\frac{(r^{\rm Sch}_p)^2}{b^2_p}
		  - 1 + \frac{2m}{r^{\rm Sch}_p}\right],
  \quad p = 1, 2, 3,
  \label{eq:trajectorySch1}
\end{align}
where $b_p$ denotes an impact parameter of null geodesic $\Gamma_p$.
Changing the variable $u^{\rm Sch}_p = 1/r^{\rm Sch}_p$,
Eq. (\ref{eq:trajectorySch1}) becomes
\begin{align}
 \left(\frac{du^{\rm Sch}_p}{d\phi}\right)^2 = \frac{1}{b^2_p}
  - (u^{\rm Sch}_p)^2 + 2m(u^{\rm Sch}_p)^3.
  \label{eq:trajectorySch2}
\end{align}
We derive the trajectories of null geodesics $\Gamma_1$, $\Gamma_2$, 
and $\Gamma_3$, three of which configure the triangle on the optical 
reference geometry ${\cal M}^{\rm opt}$ 
(see Figure \ref{fig:arakida-fig2}). 
In these cases, the zeroth-order solutions are
\begin{align}
 u^{\rm Sch, 0}_{1} = \frac{\sin\phi}{b_1},\quad
 u^{\rm Sch, 0}_{2} = \frac{\sin(\phi + \delta_2)}{b_2},\quad
 u^{\rm Sch, 0}_{3} = \frac{\sin(\phi - \delta_3)}{b_3}.
\end{align}
In accordance with the standard perturbation scheme, 
we express the solution $u^{\rm Sch}_p = u^{\rm Sch}_p (\phi)$ as
\begin{align}
 u^{\rm Sch}_{p} = u^{\rm Sch, 0}_{p}
  + \varepsilon u^{\rm Sch, 1}_{p} + \varepsilon^2 u^{\rm Sch, 2}_{p},
  \label{eq:usoltion}
\end{align}
where $\varepsilon$ denotes the small dimensionless expansion parameter, 
which in the Schwarzschild spacetime is $\varepsilon = m/b$.
The terms $\varepsilon u^{\rm Sch, 1}_{p}$ and 
$\varepsilon^2 u^{\rm Sch, 2}_{p}$ denote the first 
${\cal O}(\varepsilon)$ and second ${\cal O}(\varepsilon^2)$ order 
correction terms, respectively.
Substituting Eq (\ref{eq:usoltion}) into Eq. (\ref{eq:trajectorySch2}), 
we obtain the second-order solution with respect to $\varepsilon$ as
\begin{align}
 u^{\rm Sch}_{1} &= \frac{\sin\phi}{b_1}
  + \frac{m}{2b_1^2}(3 + \cos 2\phi)
  \nonumber\\
  &+ \frac{m^2}{16b_1^3}
  \left[
   37\sin\phi + 30(\pi - 2\phi)\cos\phi - 3\sin 3\phi
  \right] + {\cal O}(\varepsilon^3),
  \label{eq:u1_trajectory}\\
  u^{\rm Sch}_{2} &= \frac{\sin(\phi + \delta_2)}{b_2}
   + \frac{m}{2b_2^2}\left[3 + \cos 2(\phi + \delta_2)\right]
  \nonumber\\
  &+ \frac{m^2}{16b_2^3}
  \left\{
   37\sin(\phi + \delta_2) + 30[\pi - 2(\phi + \delta_2)]
   \cos(\phi + \delta_2) 
   - 3\sin 3(\phi + \delta_2)\right\}  
   + {\cal O}(\varepsilon^3),
  \label{eq:u2_trajectory}\\
  u^{\rm Sch}_{3} &= \frac{\sin(\phi - \delta_3)}{b_3}
  + \frac{m}{2b_3^2}\left[3 + \cos 2(\phi - \delta_3)\right]
  \nonumber\\
  &+ \frac{m^2}{16b_3^3}
  \left\{
   37\sin(\phi - \delta_3) + 30[\pi - 2(\phi - \delta_3)]
   \cos(\phi - \delta_3) 
   - 3\sin 3(\phi - \delta_3)\right\} + {\cal O}(\varepsilon^3).
  \label{eq:u3_trajectory}
\end{align}
The integration constants of Eqs. (\ref{eq:u1_trajectory}),
(\ref{eq:u2_trajectory}), and (\ref{eq:u3_trajectory})
are chosen to maximize $u$ (or minimize $r$)
at $\phi = \pi/2$, $\phi = \pi/2 - \delta_2$, and
$\phi = \pi/2 + \delta_3$, respectively:
\begin{align}
 \left.
 \frac{du^{\rm Sch}_{1}}{d\phi}
  \right|_{\phi = \pi/2} = 0,\quad
  \left.
 \frac{du^{\rm Sch}_{2}}{d\phi}
     \right|_{\phi = \pi/2 - \delta_2} = 0,\quad
 \left.
 \frac{du^{\rm Sch}_{3}}{d\phi}
  \right|_{\phi = \pi/2 + \delta_3} = 0.  
\end{align}
\subsection{Angular Formula}
We compute the total deflection angle $\alpha_{\rm Sch}$
by using the angular formula, i.e., Eq. (\ref{eq:deftotalangle2}).
First, we calculate the intersection angles $\psi$ between the null 
geodesics $\Gamma_1$, $\Gamma_2$, and $\Gamma_3$, and the radial 
null geodesics, $\Gamma_R$, $\Gamma_M$, and $\Gamma_S$, respectively.  
Substituting Eqs. (\ref{eq:trajectorySch1}), (\ref{eq:u1_trajectory}),
(\ref{eq:u2_trajectory}), and (\ref{eq:u3_trajectory}) into
Eq. (\ref{eq:psi1}) and expanding up to the second-order with respect to 
$\varepsilon$, we have
\begin{align}
 \psi_1^{\rm Sch} &= \phi + \frac{2m}{b_1}\cos\phi
  + \frac{m^2}{8b_1^2}
  \left[15(\pi - 2\phi) - \sin 2\phi\right]
  + {\cal O}(\varepsilon^3),
  \label{eq:psi_1_Sch}\\
 \psi_2^{\rm Sch} &= \phi + \delta_2
  + \frac{2m}{b_2}\cos (\phi + \delta_2)
  + \frac{m^2}{8b_2^2}
  \left[15(\pi - 2\phi - 2\delta_2)
  - \sin 2(\phi + \delta_2)\right]
  + {\cal O}(\varepsilon^3),
  \label{eq:psi_2_Sch}\\
 \psi_3^{\rm Sch} &= \phi - \delta_3 + \frac{2m}{b_3}\cos (\phi - \delta_3)
  + \frac{m^2}{8b_3^2}
  \left[15(\pi - 2\phi + 2\delta_3)
  -  \sin 2(\phi - \delta_3)\right]
  + {\cal O}(\varepsilon^3).
  \label{eq:psi_3_Sch}
\end{align}
Using Eqs. (\ref{eq:betaR}), (\ref{eq:betaM}), and (\ref{eq:betaS}),
internal angles $\beta_R$, $\beta_M$, and $\beta_S$ are
given as
\begin{align}
 \beta_R^{\rm Sch}
 &= \delta_2
  + \frac{2m}{b_2}\cos (\phi_R + \delta_2)
  - \frac{2m}{b_1}\cos\phi_R
  \nonumber\\
 &+ \frac{m^2}{8b_2^2}
  \left[15(\pi - 2\phi_R - 2\delta_2) - \sin 2(\phi_R + \delta_2)\right]
 \nonumber\\
 &- \frac{m^2}{8b_1^2}
  \left[15(\pi - 2\phi_R) - \sin 2\phi_R\right]
  + {\cal O}(\varepsilon^3),
  \label{eq:beta_R}\\
 \beta_M^{\rm Sch} 
 &= \delta_3 - \delta_2 + \pi
  + \frac{2m}{b_3}\cos (\phi_M - \delta_3)
  - \frac{2m}{b_2}\cos (\phi_M + \delta_2)
  \nonumber\\
 &+ \frac{m^2}{8b_3^2}
  \left[15(\pi - 2\phi_M + 2\delta_3) - \sin 2(\phi_M - \delta_3)\right]
  \nonumber\\
 &- \frac{m^2}{8b_2^2}
  \left[15(\pi - 2\phi_M - 2\delta_2) - \sin 2(\phi_M + \delta_2)\right]
  + {\cal O}(\varepsilon^3),
  \label{eq:beta_M}\\
 \beta_S^{\rm Sch} 
 &= - \delta_3 + \frac{2m}{b_1}\cos\phi_S
  - \frac{2m}{b_3}\cos (\phi_S - \delta_3)
  \nonumber\\
 &+ \frac{m^2}{8b_1^2}
  \left[15(\pi - 2\phi_S) - \sin 2\phi_S\right]
  \nonumber\\
 &- \frac{m^2}{8b_3^2}
  \left[15(\pi - 2\phi_S + 2\delta_3) - \sin 2(\phi_S -\delta_3)\right]
  + {\cal O}(\varepsilon^3).
  \label{eq:beta_S}
\end{align}
Because the background of the Schwarzschild spacetime is the flat 
Minkowski spacetime, one has
\begin{align}
  \sum_{p = 1}^{3}\beta^{\rm BG}_p 
  = 
  \sum_{p = 1}^{3}\beta^{\rm Min}_p
  =
  \pi.
  \label{eq:beta_Min}
\end{align}
Substituting Eqs. (\ref{eq:beta_R}), (\ref{eq:beta_M}), (\ref{eq:beta_S}),
and (\ref{eq:beta_Min}) into (\ref{eq:deftotalangle2}), we obtain 
the total deflection angle as
\begin{align}
 \alpha_{\rm Sch} &=
  \left|\sum_{p=1}^{3}(\beta^{\rm Sch}_p - \beta^{\rm Min}_p) \right|
  = 
  \pi - (\beta^{\rm Sch}_R + \beta^{\rm Sch}_M + \beta^{\rm Sch}_S)
  \nonumber\\
 &= 2m
  \left[
   \frac{\cos\phi_R - \cos\phi_S}{b_1}
   +
   \frac{\cos (\phi_M + \delta_2) - \cos (\phi_R + \delta_2)}{b_2}
   +
   \frac{\cos (\phi_S - \delta_3) - \cos (\phi_M - \delta_3)}{b_3}
  \right]
  \nonumber\\
 &-  \frac{m^2}{4}
  \left[
   \frac{\sin 2\phi_R - \sin 2\phi_S}{2b_1^2}
   + \frac{\sin 2(\phi_M + \delta_2) - \sin 2(\phi_R + \delta_2)}{2b_2^2}
   \right.\nonumber\\
 &+
   \frac{\sin 2(\phi_S - \delta_3) - \sin 2(\phi_M - \delta_3)}{2b_3^2}
 -\left.
   15\left(
   \frac{\phi_R - \phi_S}{b_1^2}
   + \frac{\phi_M - \phi_R}{b_2^2}
   + \frac{\phi_S - \phi_M}{b_3^2}
   \right)
    \right]
 \nonumber\\
 &+ {\cal O}(\varepsilon^3),
 \label{eq:alpha_Sch1}
\end{align}
where the sign of $\alpha_{\rm Sch}$ is taken to be positive.
Notably, the internal angles $\beta_p$ can be measured, 
in principle, via actual observations using a spacecraft.
\subsection{Integral Formula}
\begin{figure}[htbp]
\begin{center}
 \includegraphics[scale=0.3,clip]{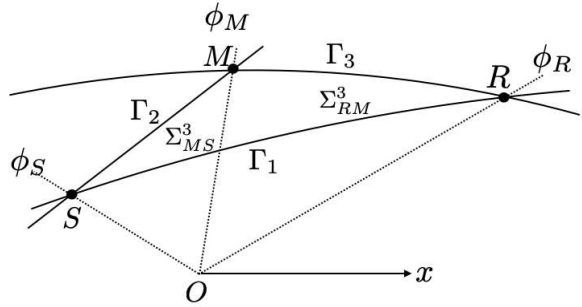}
 \caption{Dividing triangle $\Sigma^3$ into two parts.
 To integrate the areal integral of the Gaussian curvature $K$
 over the triangle $\Sigma^3$, we divide $\Sigma^3$ into 
 $\Sigma^3_{RM} (\phi_R \leq \phi \leq \phi_M)$ and 
 $\Sigma^3_{MS} (\phi_M \leq \phi \leq \phi_S)$.
 Here we assume that the angular coordinates satisfy the magnitude 
 relation, i.e., $\phi_R < \phi_M < \phi_S$.
\label{fig:arakida-fig3}}
\end{center}
\end{figure}

We show that Eq. (\ref{eq:alpha_Sch1}) can be also obtained 
using the integral formula, i.e., Eq. (\ref{eq:deftotalangle4}). 
On the optical metric Eq. (\ref{eq:optical1}), the Gaussian curvature, 
i.e., Eq. (\ref{eq:GB2}), is given as
\begin{align}
 K^{\rm Sch} = - \frac{2m}{r^3}\left(1 - \frac{3m}{2r}\right) < 0,
  \label{eq:KSch}
\end{align}
and the areal element $d\sigma^{\rm Sch}$ becomes
\begin{align}
 d\sigma^{\rm Sch}
  =
  r\left(1 - \frac{2m}{r}\right)^{-\frac{3}{2}}drd\phi.
  \label{eq:dSSch}
\end{align}
Because the background of the Schwarzschild spacetime is the flat
Minkowski spacetime, the Gaussian curvature is $K^{\rm Min} = 0$, 
and
\begin{align}
 \iint_{\Sigma^3}K^{\rm BG}d\sigma^{\rm BG}
  =
 \iint_{\Sigma^3}K^{\rm Min}d\sigma^{\rm Min} 
 = 0.
  \label{eq:KMin}
\end{align}
We divide triangle $\Sigma^3$, which is bounded by three geodesics
$\Gamma_1$, $\Gamma_2$, and $\Gamma_3$, into two parts, 
namely, $\Sigma^3_{RM}(\phi_R \leq \phi \leq \phi_M)$ and 
$\Sigma^3_{MS}(\phi_M \leq \phi \leq \phi_S)$, by assuming 
$\phi_R < \phi_M < \phi_S$ (see Figure \ref{fig:arakida-fig3}). 
Expanding up to the second order with respect to 
$\varepsilon = m/b$, the areal integral of $K^{\rm Sch}$ yields 
the total deflection angle as
\begin{align}
 \alpha_{\rm Sch} &=
  \left|
   \iint_{\Sigma^3}K^{\rm Sch}d\sigma^{\rm Sch}
   -
   \iint_{\Sigma^3}K^{\rm Min}d\sigma^{\rm Min}
	 \right|
  =
  -\iint_{\Sigma^3}K^{\rm Sch}d\sigma^{\rm Sch}
  \nonumber\\
  &= \int_{\phi_R}^{\phi_M}\int_{r^{\rm Sch}_1}^{r^{\rm Sch}_2}
  \left(
   \frac{2m}{r^2} + \frac{3m^2}{r^3} 
  \right)drd\phi
  + \int_{\phi_M}^{\phi_S}\int_{r^{\rm Sch}_1}^{r^{\rm Sch}_3}
  \left(
   \frac{2m}{r^2} + \frac{3m^2}{r^3} 
  \right)drd\phi + {\cal O}(\varepsilon^3)
  \nonumber\\
 &=
  2m
  \left[
   \frac{\cos\phi_R - \cos\phi_S}{b_1}
   +
   \frac{\cos (\phi_M + \delta_2) - \cos (\phi_R + \delta_2)}{b_2}
   +
   \frac{\cos (\phi_S - \delta_3) - \cos (\phi_M - \delta_3)}{b_3}
  \right]
  \nonumber\\
 &- \frac{m^2}{4}
  \left[
   \frac{\sin 2\phi_R - \sin 2\phi_S}{2b_1^2}
   + \frac{\sin 2(\phi_M + \delta_2) - \sin 2(\phi_R + \delta_2)}{2b_2^2}
   \right.\nonumber\\
 &+
    \frac{\sin 2(\phi_S - \delta_3) - \sin 2(\phi_M - \delta_3)}{2b_3^2}
  -\left.
   15\left(
   \frac{\phi_R - \phi_S}{b_1^2}
   + \frac{\phi_M - \phi_R}{b_2^2}
   + \frac{\phi_S - \phi_M}{b_3^2}
   \right)
    \right]
 \nonumber\\
 &+ {\cal O}(\varepsilon^3),
  \label{eq:angle_Sch2}
\end{align}
where we take the sign of $\alpha_{\rm Sch}$ 
to be positive. Additionally, $r^{\rm Sch}_1$, $r^{\rm Sch}_2$, 
and $r^{\rm Sch}_3$ are given by Eqs. (\ref{eq:u1_trajectory}), 
(\ref{eq:u2_trajectory}), and (\ref{eq:u3_trajectory}), respectively.
\subsection{Limit of Infinite Source-Observer Distance}
\begin{figure}[htbp]
\begin{center}
 \includegraphics[scale=0.3]{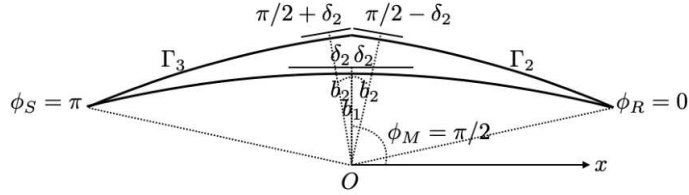}
 \caption{Symmetric triangle.
 We re-arrange triangle $\Sigma^3$ such that it is symmetrical
 with respect to $\phi = \pi/2$, and we set $b_2 = b_3$ and
 $\delta_2 = \delta_3$. The source $S$ and observer $R$ 
 are located at infinity.\label{fig:arakida-fig4}}
\end{center}
\end{figure}

\begin{figure}[htbp]
\begin{center}
 \includegraphics[scale=0.3,clip]{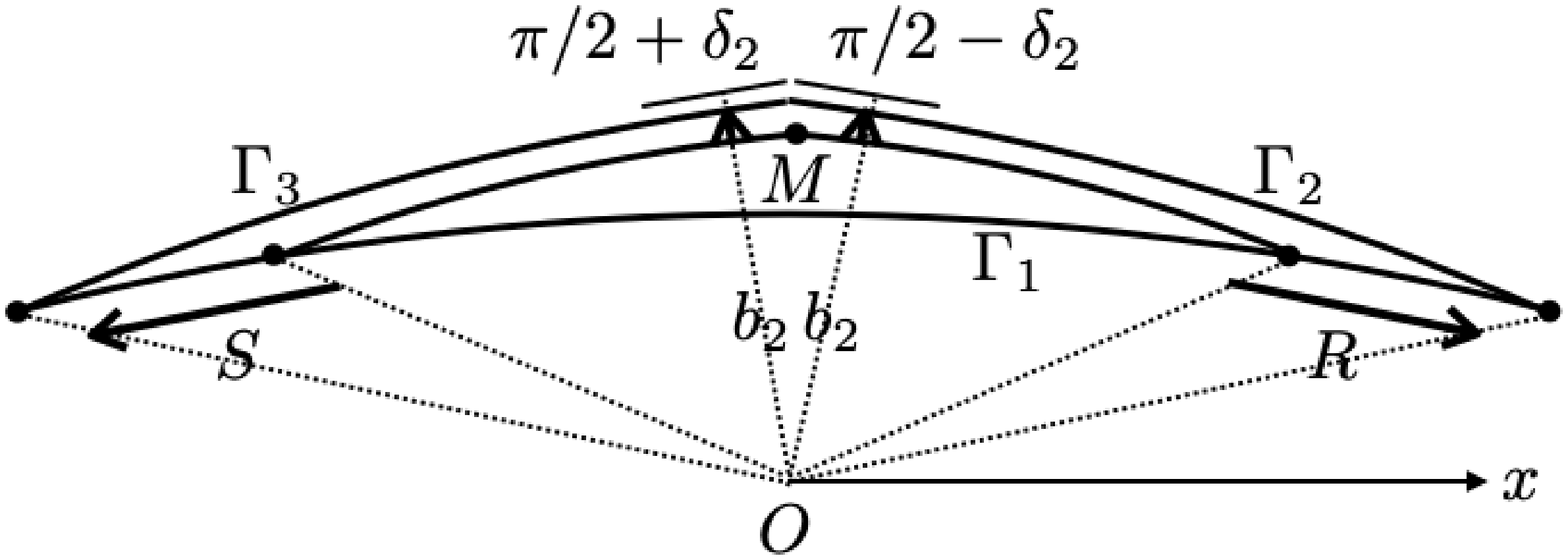}
 \caption{Relationship between two points $S$ and $R$ of triangle 
 $\Sigma^3$ and its size. This figure shows that as 
 the two points $S$ and $R$ approach infinity, the 
 impact parameter $b_2$ of $\Gamma_2$ and $\Gamma_3$  
 also becomes infinite, i.e., $b_2 \rightarrow \infty$.
 \label{fig:arakida-fig5}}
\end{center}
\end{figure}

Let us confirm that Eqs. (\ref{eq:alpha_Sch1}) and 
(\ref{eq:angle_Sch2}) can reproduce Epstein--Shapiro's 
formula \cite{epstein1980}. To this end, as shown 
in Figure \ref{fig:arakida-fig4}, we re-arrange the triangle such 
that it is symmetric with respect to
$\phi = \pi/2$; we put $\phi_M = \pi/2$, $b_2 = b_3$, and 
$\delta_3 = \delta_2$. Additionally, letting source $S$ and observer $R$
to be located at asymptotically infinite flat regions 
$\phi_S \rightarrow \pi$ and $\phi_R \rightarrow 0$, respectively, 
we obtain
\begin{align}
 \alpha^{\rm Sch}
  \rightarrow 
  4m\left(
     \frac{1}{b_1} - \frac{\sin\delta_2 + \cos\delta_2}{b_2}
    \right)
  + \frac{m^2}{4}
  \left(
   \frac{15\pi}{b_1^2} + \frac{2\sin 2\delta_2 - 15\pi}{b^2_2}
  \right)
  + {\cal O}(\varepsilon^3),
  \label{eq:Schlimit1}
\end{align}
where the following two terms:
\begin{align*}
 -4m \frac{\sin\delta_2 + \cos\delta_2}{b_2},\quad
 \frac{m^2}{4}
 \frac{2\sin 2\delta_2 - 15\pi}{b^2_2},
\end{align*}
are newly appeared.
However, as the two points $S$ and $R$ approach infinity, the 
impact parameter $b_2$ of $\Gamma_2$ and $\Gamma_3$  
becomes infinite, i.e., $b_2 \rightarrow \infty$ 
(see Figure \ref{fig:arakida-fig5}). 
Accordingly, Eq. (\ref{eq:Schlimit1}) results in
\begin{align}
 \alpha^{\rm Sch} \rightarrow \frac{4m}{b_1} + \frac{15\pi m^2}{4b_1^2} 
  + {\cal O}(\varepsilon^3).
  \label{eq:Schlimit2}
\end{align}
Notably, null geodesics $\Gamma_2$ and $\Gamma_3$ are 
not the asymptotes of null geodesic $\Gamma_1$.
\section{Total Deflection Angle in the Schwarzschild--de Sitter 
Spacetime\label{sec:Schwarzschild-deSitter}}
Let us derive the expression for the total deflection angle
and investigate the contribution of the cosmological constant 
$\Lambda$ to the total deflection angle in the Schwarzschild--de 
Sitter/Kottler spacetime \cite{kottler1918}, which is characterized as
\begin{align}
 f^{\rm SdS}(r) = 1 - \frac{2m}{r} - \frac{\Lambda}{3}r^2,
  \label{eq:fSdS}
\end{align}
where $\Lambda$ denotes the cosmological constant.
\subsection{Light Trajectory}
The first-order differential equation for null geodesics becomes
\begin{align}
 \left(\frac{dr^{\rm SdS}_p}{d\phi}\right)^2
  = (r^{\rm SdS}_p)^2\left[(r^{\rm SdS}_p)^2
	\left(
	 \frac{1}{b^2_p} + \frac{\Lambda}{3}
	\right)
	- 1 + \frac{2m}{r^{\rm SdS}_p}\right].
  \label{eq:trajectorySdS1}
\end{align}
Here, let us introduce another constant $B$ as 
\begin{align}
 \frac{1}{B^2_p} \equiv \frac{1}{b^2_p} + \frac{\Lambda}{3},
  \label{eq:B}
\end{align}
and we rewrite Eq. (\ref{eq:trajectorySdS1}) as
\begin{align}
 \left(\frac{dr^{\rm SdS}_p}{d\phi}\right)^2
  = (r^{\rm SdS}_p)^2\left[\frac{(r^{\rm SdS}_p)^2}{B^2_p}
		  - 1 + \frac{2m}{r^{\rm SdS}_p}\right],
  \label{eq:trajectorySdS2}
\end{align}
which is of the same form as Eq. (\ref{eq:trajectorySch1}).
Therefore, up to the second order in $\varepsilon = m/B$
instead of $\varepsilon = m/b$, 
the equation of the light trajectory in the Schwarzschild--de Sitter 
spacetime can be expressed as of the same form of Eqs. 
(\ref{eq:u1_trajectory}), (\ref{eq:u2_trajectory}), and 
(\ref{eq:u3_trajectory}), 
\begin{align}
 u_1^{\rm SdS} &= \frac{\sin\phi}{B_1}
  + \frac{m}{2B_1^2}(3 + \cos 2\phi)
  \nonumber\\
  &+ \frac{m^2}{16B_1^3}
  \left[
   37\sin\phi + 30(\pi - 2\phi)\cos\phi - 3\sin 3\phi
  \right] + {\cal O}(\varepsilon^3),
  \label{eq:u1_trajectorySdS}\\
  u_2^{\rm SdS} &= \frac{\sin(\phi + \delta_2)}{B_2}
   + \frac{m}{2B_2^2}\left[3 + \cos 2(\phi + \delta_2)\right]
  \nonumber\\
  &+ \frac{m^2}{16B_2^3}
  \left\{
   37\sin(\phi + \delta_2) + 30[\pi - 2(\phi + \delta_2)]
   \cos(\phi + \delta_2) 
   -3\sin 3(\phi + \delta_2)\right\}
 + {\cal O}(\varepsilon^3),
  \label{eq:u2_trajectorySdS}\\
  u_3^{\rm SdS} &= \frac{\sin(\phi - \delta_3)}{B_3}
  + \frac{m}{2B_3^2}\left[3 + \cos 2(\phi - \delta_3)\right]
  \nonumber\\
  &+ \frac{m^2}{16B_3^3}
  \left\{
   37\sin(\phi - \delta_3) + 30[\pi - 2(\phi - \delta_3)]
   \cos(\phi - \delta_3) 
   -3\sin 3(\phi - \delta_3)\right\}
 + {\cal O}(\varepsilon^3).
  \label{eq:u3_trajectorySdS}
\end{align}
The equation of light trajectory in the
Schwarzschild--de Sitter spacetime does not depend on the cosmological 
constant $\Lambda$ and impact parameter $b$. 
Using Eq. (\ref{eq:trajectorySdS2}) and the condition
\begin{align}
 \left.
  \frac{dr}{d\phi}
  \right|_{r = r_0} = 0,
\end{align}
we have following relation:
\begin{align}
 \frac{1}{B^2} = \frac{1}{b^2} + \frac{\Lambda}{3}
  = \frac{1}{r_0^2} - \frac{2m}{r_0^3},
\end{align}
where $r_0$ denotes the radial coordinate value at the closest
approach of the light ray. In principle, $r_0$ can be obtained via 
actual measurements, as the circumference radius $\ell_0 = 2\pi r_0$. 
Therefore, $B$ is calculated without knowing the value of $\Lambda$ and $b$.
However, as will be discussed below, this does not mean that the total 
deflection angle is also independent of $\Lambda$.
\subsection{Background of the Schwarzschild--de Sitter
Spacetime}
Before discussing the total deflection angle, 
let us debate the background of the Schwarzschild--de Sitter 
spacetime and derive some relations. 

In the case of the Schwarzschild--de Sitter spacetime,
the background should be regarded as the de Sitter spacetime 
instead of the Minkowski spacetime. This is because we had, 
in advance, assumed the existence of non-zero cosmological 
constant $\Lambda$, which is not an integration constant as 
mass $m$. In fact, the action
\begin{align}
 {\cal S} = \int
  \left[
   \frac{c^4}{16\pi G}(R - 2\Lambda) + {\cal L}_{M}
  \right]\sqrt{-g}d^4x 
\end{align}
and the field equation
\begin{align}
 R_{\mu\nu} - \frac{1}{2}g_{\mu\nu}R + \Lambda g_{\mu\nu}
  = \frac{8\pi G}{c^4}T_{\mu\nu}
\end{align}
explicitly include the cosmological constant $\Lambda$, where
$g = \det(g_{\mu\nu})$; the term ${\cal L}_{M}$ denotes the Lagrangian
for the matter field; $R_{\mu\nu}$ and $R$ denote the Ricci tensor
and Ricci scalar, respectively; $T_{\mu\nu}$ denotes the energy-momentum 
tensor. Hence, the spacetime cannot reduce to the Minkowski spacetime.

The de Sitter spacetime is characterized by
\begin{align}
 f^{\rm dS}(r) = 1 - \frac{\Lambda}{3}r^2,
  \label{eq:fdS}
\end{align}
and the differential equation of a light ray on the optical
reference geometry ${\cal M}^{\rm opt}$ becomes
\begin{align}
 \left(\frac{dr^{\rm dS}_p}{d\phi}\right)^2 =
  (r^{\rm dS}_p)^2
  \left[
   \left(\frac{1}{b^2_p} + \frac{\Lambda}{3}\right)(r^{\rm dS}_p)^2 - 1
       \right].
  \label{eq:dSdiffeq1}
\end{align}
Using Eq. (\ref{eq:B}), the light trajectories in the 
de Sitter spacetime are
\begin{align}
 u_{1}^{\rm dS} &= \frac{\sin\phi}{B_1},
  \label{eq:u1_trajectorydS}\\
 u_{2}^{\rm dS} &= \frac{\sin (\phi + \delta_2)}{B_2},
  \label{eq:u2_trajectorydS}\\
 u_{3}^{\rm dS} &= \frac{\sin (\phi - \delta_3)}{B_3}.
  \label{eq:u3_trajectorydS}
\end{align}
Substituting Eqs. (\ref{eq:fdS}), (\ref{eq:u1_trajectorydS}),
(\ref{eq:u2_trajectorydS}), and (\ref{eq:u3_trajectorydS}) into  
Eq. (\ref{eq:psi1}), the intersection angles $\psi$ between
three null geodesics $\Gamma_p ~ (p = 1, 2, 3)$ and 
radial null geodesics $\Gamma_R$, $\Gamma_M$, and $\Gamma_S$, respectively,  
are computed as
\begin{align}
 \psi^{\rm dS}_1 
  &=
  \arctan \left(\sqrt{1 - \frac{\Lambda B^2_1}
	   {3}\csc^2 \phi}\tan \phi\right)
  = \arccos\frac{\cos\phi}{\sqrt{1 - \frac{\Lambda B_1^2}{3}}},
  \label{eq:psi_1_dS}\\
 \psi^{\rm dS}_2 
  &= 
  \arctan \left(\sqrt{1 - \frac{\Lambda B^2_2}
	   {3}\csc^2 (\phi + \delta_2)}\tan (\phi + \delta_2)\right)
 = \arccos\frac{\cos(\phi + \delta_2)}{\sqrt{1 - \frac{\Lambda B_2^2}{3}}},
  \label{eq:psi_2_dS}\\
  \psi^{\rm dS}_3 
   &= 
  \arctan \left(\sqrt{1 - \frac{\Lambda B^2_3}
	   {3}\csc^2 (\phi - \delta_3)}\tan (\phi - \delta_3)\right)
  = \arccos\frac{\cos(\phi - \delta_3)}{\sqrt{1 - \frac{\Lambda B_3^2}{3}}},
  \label{eq:psi_3_dS}
\end{align}
where we used the following inverse trigonometric function:
\begin{align}
 \arctan \phi = \arccos \frac{1}{\sqrt{1 + \phi^2}}.
\end{align}
Using Eqs. (\ref{eq:betaR}), (\ref{eq:betaM}), and (\ref{eq:betaS}),
the sum of the internal angles, $\beta^{\rm dS}_p$, is given by
\begin{align}
 \sum_{p = 1}^{3}\beta^{\rm dS}_p
  &= \pi\nonumber\\
 &+
  \arccos\frac{\cos \phi_R}
  {\sqrt{1 - \frac{\Lambda B^2_1}{3}}}
  -
  \arccos\frac{\cos \phi_S}
  {\sqrt{1 - \frac{\Lambda B^2_1}{3}}} 
  \nonumber\\
 &+ \arccos\frac{\cos (\phi_M + \delta_2)}
  {\sqrt{1 - \frac{\Lambda B^2_2}{3}}}
  - 
  \arccos\frac{\cos (\phi_R + \delta_2)}
  {\sqrt{1 - \frac{\Lambda B^2_2}{3}}}
  \nonumber\\
 &+ \arccos\frac{\cos (\phi_S - \delta_3)}
  {\sqrt{1 - \frac{\Lambda B^2_3}{3}}}
  - 
  \arccos\frac{\cos (\phi_M - \delta_3)}
  {\sqrt{1 - \frac{\Lambda B^2_3}{3}}}.
  \label{eq:angle_dS1}
\end{align}
Notably, the sum of the internal angles of the triangle differs 
from $\pi$ because of the existence of the cosmological 
constant $\Lambda$.

On the other hand, the Gaussian curvature and areal element are 
given by
\begin{align}
 K^{\rm dS} = -\frac{\Lambda}{3} < 0,\quad
  d\sigma^{\rm dS}
  = r\left(1 - \frac{\Lambda}{3}r^2\right)^{-\frac{3}{2}}
  drd\phi.
 \label{eq:KdS}
\end{align}
As in the case of the Schwarzschild spacetime, we divide 
triangle $\Sigma^3$ into 
$\Sigma^3_{RM}(\phi_R \leq \phi \leq \phi_M)$ and 
$\Sigma^3_{MS}(\phi_M \leq \phi \leq \phi_S)$ 
(see Figure \ref{fig:arakida-fig3}).
Subsequently, using Eq. (\ref{eq:KdS}), 
the areal integral of $K^{\rm dS}$ becomes
\begin{align}
 -\iint_{\Sigma^3}K^{\rm dS}d\sigma^{\rm dS}
  &= \int^{\phi_M}_{\phi_R}\int^{r_2^{\rm dS}}_{r_1^{\rm dS}}
  \frac{\Lambda}{3}r\left(1 - \frac{\Lambda}{3}r^2\right)^{-\frac{3}{2}}
  drd\phi
  +
 \int^{\phi_S}_{\phi_M}\int^{r_3^{\rm dS}}_{r_1^{\rm dS}}
  \frac{\Lambda}{3}r\left(1 - \frac{\Lambda}{3}r^2\right)^{-\frac{3}{2}}
  drd\phi
  \nonumber\\
 &=
   \int^{\phi_M}_{\phi_R}
   \left[
    \frac{\sin (\phi + \delta_2)}
    {\sqrt{\sin^2 (\phi + \delta_2) - \frac{\Lambda B^2_2}{3}}}
    -
    \frac{\sin \phi}
    {\sqrt{\sin^2 \phi - \frac{\Lambda B^2_1}{3}}}
   \right]
  d\phi
  \nonumber\\
 &+
 \int^{\phi_S}_{\phi_M}
   \left[
    \frac{\sin (\phi - \delta_3)}
    {\sqrt{\sin^2 (\phi - \delta_3) - \frac{\Lambda B^2_3}{3}}}
    - 
    \frac{\sin \phi}
    {\sqrt{\sin^2 \phi - \frac{\Lambda B^2_1}{3}}}
   \right]
  d\phi
  \nonumber\\
 &=
  \arccos\frac{\cos \phi_S}
  {\sqrt{1 - \frac{\Lambda B^2_1}{3}}} 
  -
  \arccos\frac{\cos \phi_R}
  {\sqrt{1 - \frac{\Lambda B^2_1}{3}}}
  \nonumber\\
 &+ 
  \arccos\frac{\cos (\phi_R + \delta_2)}
  {\sqrt{1 - \frac{\Lambda B^2_2}{3}}}
  -
  \arccos\frac{\cos (\phi_M + \delta_2)}
  {\sqrt{1 - \frac{\Lambda B^2_2}{3}}}
  \nonumber\\
 &+ 
  \arccos\frac{\cos (\phi_M - \delta_3)}
  {\sqrt{1 - \frac{\Lambda B^2_3}{3}}},
  -
  \arccos\frac{\cos (\phi_S - \delta_3)}
  {\sqrt{1 - \frac{\Lambda B^2_3}{3}}}   
  \label{eq:angle_dS2}
\end{align}
where we used the following pseudo-elliptic integral, 
which is provided on p. 205 in \cite{gr2007};
\begin{align}
 \int \frac{\sin\phi}{\sqrt{a^2 \sin^2\phi - 1}}d\phi
  = - \frac{1}{a}\arcsin \frac{a \cos \phi}{\sqrt{a^2 - 1}},
  \quad a^2 > 1,
\end{align}
and the inverse trigonometric function:
$\arcsin \phi + \arccos \phi = \pi/2$.
Eqs. (\ref{eq:angle_dS1}) and (\ref{eq:angle_dS2}) are 
calculated without any approximation with respect to $\Lambda$.

Before concluding this subsection, we give some comments.
The sum of the internal angles of the triangle in the de Sitter 
spacetime differs from $\pi$ because of the non-zero Gaussian curvature 
$K^{\rm dS} = -\Lambda/3$. Accordingly, the
value of the cosmological constant $\Lambda$ may be obtained 
by taking the difference 
between $\sum_{p = 1}^{3}\beta^{\rm dS}_p$ and $\pi$.
However, as indicated previously, if we consider the existence 
of non-zero cosmological constant in the context of the
Schwarzschild--de Sitter spacetime, then the de Sitter spacetime should 
be adopted as the background instead of the Minkowski spacetime. 
In fact, the Schwarzschild--de Sitter spacetime can be considered 
the distorted de Sitter spacetime because of mass $m$ 
(see section 14.4 in \cite{rindler2006})
\subsection{Angular Formula}
First, we obtain angle $\psi$. 
We wish to express $\psi^{\rm SdS}$ as 
\begin{align}
 \psi_p^{\rm SdS} = \psi_p^{\rm dS} + 
  {\cal O}(m, m^2, m\Lambda) ~ \mbox{terms},
  \label{eq:SdScorrectionterms1}
\end{align}
where $\psi_p^{\rm dS} ~ (p = 1, 2, 3)$ are given by 
Eqs. (\ref{eq:psi_1_dS}), (\ref{eq:psi_2_dS}), and (\ref{eq:psi_3_dS}),
respectively. 

To calculate the correction term ${\cal O}(m, m^2, m\Lambda)$, 
we first set the small dimensionless expansion parameter $\varepsilon$ 
as $\varepsilon = m/B$. Subsequently, using Eqs. (\ref{eq:fSdS}), 
(\ref{eq:trajectorySdS2}), (\ref{eq:u1_trajectorySdS}), 
(\ref{eq:u2_trajectorySdS}), and (\ref{eq:u3_trajectorySdS}) 
we expand Eq. (\ref{eq:psi1}) up to the second order in 
$\varepsilon = m/B$ and obtain the following approximate expression:
\begin{align}
 \psi^{\rm SdS}_1 &= 
  \arccos\frac{\cos \phi}{\sqrt{1 - \frac{\Lambda B_1^2}{3}}}
  + 
  \frac{2m\cos\phi}{B_1\sqrt{1 - \frac{\Lambda B^2_1}{3}\csc^2\phi}}
  \nonumber\\
  &+
   \frac{m^2}
   {16B^2_1 \left(1 - \frac{\Lambda B^2_1}{3}\csc^2\phi\right)^{\frac{3}{2}}}
   \{2[15(\pi - 2\phi) - \sin 2\phi] 
   \nonumber\\
 &- \Lambda B^2_1\csc^2\phi
   (30\pi - 60\phi + 23\cot\phi + 9\cos 3\phi\csc\phi)\}
   +
   {\cal O}(\varepsilon^3),
   \label{eq:psi_0_SdS1}\\
 \psi^{\rm SdS}_2 &=
  \arccos\frac{\cos (\phi + \delta_2)}{\sqrt{1 - \frac{\Lambda B_2^2}{3}}}
  + 
  \frac{2m\cos(\phi + 
  \delta_2)}{B_2\sqrt{1 - \frac{\Lambda B^2_2}{3}\csc^2(\phi + \delta_2)}}
  \nonumber\\
  &+
   \frac{m^2}{16B^2_2\left[1 - 
	\frac{\Lambda B^2_2}{3}\csc^2(\phi + \delta_2)\right]^{\frac{3}{2}}}
   \left(2\{15[\pi - 2(\phi + \delta_2)] - \sin 2(\phi + \delta_2)\}
    \right.\nonumber\\
 &-\left. 
     \Lambda B^2_2\csc^2(\phi + \delta_2)
   [30\pi - 60(\phi + \delta_2)\right.
 \nonumber\\
 &+\left. 23\cot(\phi + \delta_2) 
   + 9\cos 3(\phi + \delta_2)\csc(\phi + \delta_2)]\right)
   +
   {\cal O}(\varepsilon^3),
   \label{eq:psi_0_SdS2}\\
 \psi^{\rm SdS}_3 &=
  \arccos\frac{\cos (\phi - \delta_3)}{\sqrt{1 - \frac{\Lambda B_3^2}{3}}}
  + 
  \frac{2m\cos(\phi - \delta_3)}
  {B_3\sqrt{1 - \frac{\Lambda B^2_3}{3}\csc^2(\phi - \delta_3)}}
  \nonumber\\
  &+
   \frac{m^2}{16B^2_3\left[1 - 
	\frac{\Lambda B^2_3}{3}\csc^2(\phi - \delta_3)\right]^{\frac{3}{2}}}
      \left(2\{15[\pi - 2(\phi - \delta_3)] - \sin 2(\phi - \delta_3)\}
    \right.\nonumber\\
 &- \left. 
     \Lambda B^2_3\csc^2(\phi - \delta_3)
     [30\pi - 60(\phi - \delta_3)\right.
 \nonumber\\
 &+ \left. 23\cot(\phi - \delta_3) + 
     9\cos 3(\phi - \delta_3)\csc(\phi - \delta_3)]\right)
   +
    {\cal O}(\varepsilon^3).
    \label{eq:psi_0_SdS3}
\end{align}
Note that at this stage, we do not adopt $\varepsilon = \Lambda B^2$
as the small dimensionless expansion parameter. The first terms 
in the first lines of Eqs. (\ref{eq:psi_0_SdS1}), (\ref{eq:psi_0_SdS2}),
and (\ref{eq:psi_0_SdS3}) are equal to 
$\psi^{\rm dS}_1$, $\psi^{\rm dS}_2$, and $\psi^{\rm dS}_3$,
respectively (see Eqs. (\ref{eq:psi_1_dS}), (\ref{eq:psi_2_dS}),
and (\ref{eq:psi_3_dS})).

Next, expanding ${\cal O}(m)$ and ${\cal O}(m^2)$ terms in 
Eqs. (\ref{eq:psi_0_SdS1}), (\ref{eq:psi_0_SdS2}), and
(\ref{eq:psi_0_SdS3}) with respect to $\varepsilon = \Lambda B^2$ 
and the remaining ${\cal O}(m/B, (m/B)^2, (m/B)\cdot \Lambda B^2)$ terms, 
we have
\begin{align}
 \psi^{\rm SdS}_1 &= \psi^{\rm dS}_1 + \frac{2m}{B_1}\cos\phi
 + \frac{m^2}{8B_1^2}[15(\pi - 2\phi) - \sin 2\phi]
 \nonumber\\
  &+ \frac{\Lambda m B_1}{3}\cot\phi\csc\phi
  + {\cal O}(\varepsilon^3),
  \label{eq:psi_1_SdS}\\
  \psi^{\rm SdS}_2 &= \psi^{\rm dS}_2
   + \frac{2m}{B_2}\cos 2(\phi + \delta_2)
   + \frac{m^2}{8B_2^2}\{15[\pi - 2(\phi + \delta_2)]
  - \sin 2(\phi + \delta_2)\}
  \nonumber\\
  &+ \frac{\Lambda m B_2}{3}\cot(\phi + \delta_2)\csc(\phi + \delta_2)
  + {\cal O}(\varepsilon^3),
  \label{eq:psi_2_SdS}\\
 \psi^{\rm SdS}_3 &= \psi^{\rm dS}_3 
  + \frac{2m}{B_3}\cos(\phi - \delta_3)
 + \frac{m^2}{8B_3^2}\{15[\pi - 2(\phi - \delta_3)]
  - \sin 2(\phi - \delta_3)\}
  \nonumber\\
  &+ \frac{\Lambda m B_3}{3}\cot(\phi - \delta_3)\csc(\phi - \delta_3)
  + {\cal O}(\varepsilon^3),
  \label{eq:psi_3_SdS}
\end{align}
where the residual terms ${\cal O}(\varepsilon^3)$ are 
${\cal O}((m/B)^3, (m/B)^2\cdot\Lambda B^2, (m/B)\cdot (\Lambda B^2)^2)$.
Notably, $\psi^{\rm SdS}_1$, $\psi^{\rm SdS}_2$, and 
$\psi^{\rm SdS}_3$ comprise the terms characterized by the 
cosmological constant $\Lambda$, namely $\psi^{\rm dS}_1$, 
$\psi^{\rm dS}_2$, and 
$\psi^{\rm dS}_3$. However as we will see below,
the terms $\psi^{\rm dS}_1$, $\psi^{\rm dS}_2$, and $\psi^{\rm dS}_3$ 
are eliminated from the expression of the total deflection
angle $\alpha_{\rm SdS}$.

Using Eqs. (\ref{eq:deftotalangle2}), (\ref{eq:betaR}), (\ref{eq:betaM}),
(\ref{eq:betaS}), (\ref{eq:angle_dS1}), (\ref{eq:psi_1_SdS}), 
(\ref{eq:psi_2_SdS}), and (\ref{eq:psi_3_SdS}), 
the total deflection angle becomes
\begin{align}
  \alpha_{\rm SdS} &=
  \left|\sum_{p=1}^{3}(\beta^{\rm SdS}_p - \beta^{\rm dS}_p) \right|
  = 
  (\beta_R^{\rm dS} + \beta_M^{\rm dS} + \beta_S^{\rm dS})
  - 
  (\beta_R^{\rm SdS} + \beta_M^{\rm SdS} + \beta_S^{\rm SdS})
  \nonumber\\
 &=
  2m
  \left[
   \frac{\cos\phi_R - \cos\phi_S}{B_1}
   +
   \frac{\cos (\phi_M + \delta_2) - \cos (\phi_R + \delta_2)}{B_2}
   +
   \frac{\cos (\phi_S - \delta_3) - \cos (\phi_M - \delta_3)}{B_3}
  \right]
  \nonumber\\
 &- \frac{m^2}{4}
  \left[
   \frac{\sin 2\phi_R - \sin 2\phi_S}{2B_1^2}
   + \frac{\sin 2(\phi_M + \delta_2) - \sin 2(\phi_R + \delta_2)}{2B_2^2}
   \right.\nonumber\\
 &-
   \frac{\sin 2(\phi_M - \delta_3) - \sin 2(\phi_S - \delta_3)}{2B_3^2}
   -\left.
   15\left(
   \frac{\phi_R - \phi_S}{B_1^2}
   + \frac{\phi_M - \phi_R}{B_2^2}
   - \frac{\phi_M - \phi_S}{B_3^2}
   \right)
    \right]\nonumber\\
 &+
  \frac{\Lambda B_1 m}{3}\cot\phi_R\csc\phi_R
  -
  \frac{\Lambda B_1 m}{3}\cot\phi_S\csc\phi_S.
  \nonumber\\
 &+
  \frac{\Lambda B_2 m}{3}\cot(\phi_M + \delta_2)\csc(\phi_M + \delta_2)
 -
  \frac{\Lambda B_2 m}{3}\cot(\phi_R + \delta_2)\csc(\phi_R + \delta_2)
  \nonumber\\
 &+
  \frac{\Lambda B_3 m}{3}\cot(\phi_S - \delta_3)\csc(\phi_S - \delta_3)
 -
  \frac{\Lambda B_3 m}{3}\cot(\phi_M - \delta_3)\csc(\phi_M - \delta_3)
 \nonumber\\
  &+
  {\cal O}(\varepsilon^3).
  \label{eq:defangleSdS1}
\end{align}
In Eq. (\ref{eq:defangleSdS1}), the second to sixth lines correspond to
the Schwarzschild-like part and the seventh to twelfth lines are order 
${\cal O}(m\Lambda)$ terms due to the cosmological constant $\Lambda$.
The contribution of the cosmological constant $\Lambda$
appears as ${\cal O}(\Lambda m)$ instead of 
${\cal O}(\Lambda/m)$. Because $\psi^{\rm SdS}_p$ can be expressed 
as Eq. (\ref{eq:SdScorrectionterms1}), $\psi^{\rm dS}_1$, 
$\psi^{\rm dS}_2$, and $\psi^{\rm dS}_3$ are completely eliminated from 
in $\alpha_{\rm SdS}$. Therefore, 
$\alpha_{\rm SdS}$ does not include the terms described solely
by the cosmological constant $\Lambda$.
\subsection{Integral Formula}
We compute the total deflection angle $\alpha_{\rm SdS}$
using the integral formula.
As in the case of the angular formula, we represent the areal 
integral of the Gaussian curvature $K^{\rm SdS}$ as
\begin{align}
 -\iint_{\Sigma^3}K^{\rm SdS}d\sigma^{\rm SdS}
  =
  - \iint_{\Sigma^3}K^{\rm dS}d\sigma^{\rm dS}
  + {\cal O}(m, m^2, m\Lambda) ~ \mbox{terms},
  \label{eq:SdScorrectionterms2}
\end{align}
where the areal integral of $K^{\rm dS}$ is given by Eq. 
(\ref{eq:angle_dS2}).

The Gaussian curvature $K^{\rm SdS}$ in the Schwarzschild--de
Sitter spacetime becomes
\begin{align}
 K^{\rm SdS} = - \frac{2m}{r^3}
  \left(1 - \frac{3m}{2r} + \frac{\Lambda}{6m}r^3
   - \Lambda r^2\right) < 0,
  \label{eq:KSdS}
\end{align}
and the areal element $d\sigma^{\rm SdS}$ is given by
\begin{align}
  d\sigma^{\rm SdS} = r
  \left(1 - \frac{2m}{r} - \frac{\Lambda}{3}r^2\right)^{-\frac{3}{2}}
  drd\phi,
  \label{eq:dSSdS}
\end{align}
Similar to the same way in which Eqs. (\ref{eq:angle_Sch2}) and 
(\ref{eq:angle_dS1}) are integrated,
we construct triangle $\Sigma^3$, which is bounded by three geodesics
$\Gamma_1$, $\Gamma_2$, and $\Gamma_3$, and divide the triangle
$\Sigma^3$ into
$\Sigma^3_{RM}(\phi_R \leq \phi \leq \phi_M)$ and 
$\Sigma^3_{MS}(\phi_M \leq \phi \leq \phi_S)$ (see 
Figure \ref{fig:arakida-fig3}).

Before integrating the areal integral of the Gaussian curvature over
the triangle $\Sigma^3$, we approximate the integrand of the areal 
integral up to the second order in $\varepsilon = m/B$ as follows:
\begin{align}
 - \iint_{\Sigma^3}K^{\rm SdS}d\sigma^{\rm SdS}
  &=
  \iint_{\Sigma^3}\frac{\Lambda}{3}r 
  \left(1 - \frac{\Lambda}{3}r^2\right)^{-\frac{3}{2}}drd\phi
  \nonumber\\
  &+
  \iint_{\Sigma^3}
  \left\{
  \frac{m[6 - \Lambda r^2 (5 - 2\Lambda r^2)]}
  {3r^2 \left(1 - \frac{\Lambda}{3}r^2\right)^{\frac{5}{2}}}
  +
  \frac{m^2[18 - \Lambda r^2 (21 - 10 \Lambda r^2)]}
  {6 r^3 \left(1 - \frac{\Lambda}{3}r^2\right)^{\frac{7}{2}}}
  \right\}drd\phi
  \nonumber\\
  &+
  {\cal O}(\varepsilon^3).
  \label{eq:area_SdS1}
\end{align}
Note that at this stage in Eq. (\ref{eq:area_SdS1}), 
$\varepsilon = \Lambda B^2$ is not treated as a small dimensionless 
expansion parameter.

The first term in the right-hand side of Eq. (\ref{eq:area_SdS1}) 
has the same form as that of Eq. (\ref{eq:angle_dS2}) in the de Sitter 
spacetime. 
However, the light trajectory $r$ is $r^{\rm SdS}_p$ instead of 
$r^{\rm dS}_p$. Let us compute the first term of 
Eq. (\ref{eq:area_SdS1}). Substituting 
Eqs. (\ref{eq:u1_trajectorySdS}), (\ref{eq:u2_trajectorySdS}), 
and (\ref{eq:u3_trajectorySdS}) into the first term of 
Eq. (\ref{eq:area_SdS1}), integrating over $r$, and remaining 
$\varepsilon = m/B$ order terms,  we have
\begin{align}
 &\quad~
 \iint_{\Sigma^3}\frac{\Lambda}{3}r 
  \left(1 - \frac{\Lambda}{3}r^2\right)^{-\frac{3}{2}}drd\phi
  \nonumber\\
  &=
  \int^{\phi_M}_{\phi_R}\int^{r^{\rm SdS}_2}_{r^{\rm SdS}_1}
  \frac{\Lambda}{3}r 
  \left(1 - \frac{\Lambda}{3}r^2\right)^{-\frac{3}{2}}drd\phi
  +
  \int^{\phi_S}_{\phi_M}\int^{r^{\rm SdS}_3}_{r^{\rm SdS}_1}
  \frac{\Lambda}{3}r 
  \left(1 - \frac{\Lambda}{3}r^2\right)^{-\frac{3}{2}}drd\phi
  \nonumber\\
  &=
  \int^{\phi_M}_{\phi_R}
  \left[
  \left(1 - \frac{\Lambda}{3}r^2\right)^{-\frac{1}{2}}
  \right]^{r^{\rm SdS}_2}_{r^{\rm SdS}_1}
  d\phi
  +
  \int^{\phi_S}_{\phi_M}
  \left[
   \left(1 - \frac{\Lambda}{3}r^2\right)^{-\frac{1}{2}}
       \right]^{r^{\rm SdS}_3}_{r^{\rm SdS}_1}
  d\phi
  \nonumber\\
 &=
  - \iint_{\Sigma^3}K^{\rm dS} d\sigma^{\rm dS}
  \nonumber\\
 &-
  \int^{\phi_M}_{\phi_R}
  \left\{
   \frac{\Lambda B_2 m [3 + \cos 2(\phi + \delta_2)]
   \csc^3 (\phi + \delta_2)}
   {6\left(1 - \frac{\Lambda}{3}B^2_2
      \csc^2 (\phi + \delta_2)\right)^{\frac{3}{2}}}
   -
   \frac{\Lambda B_1 m (3 + \cos 2 \phi)
   \csc^3 \phi}
   {6\left(1 - \frac{\Lambda}{3}B^2_1
      \csc^2 \phi\right)^{\frac{3}{2}}}
  \right\}d\phi
  \nonumber\\
  &-
  \int^{\phi_S}_{\phi_M}
  \left\{
   \frac{\Lambda B_3 m [3 + \cos 2(\phi - \delta_3)]
   \csc^3 (\phi - \delta_3)}
   {6\left(1 - \frac{\Lambda}{3}B^2_3
      \csc^2 (\phi - \delta_3)\right)^{\frac{3}{2}}}
   -
   \frac{\Lambda B_1 m (3 + \cos 2 \phi)
   \csc^3 \phi}
   {6\left(1 - \frac{\Lambda}{3}B^2_1
      \csc^2 \phi\right)^{\frac{3}{2}}}
  \right\}d\phi
  + {\cal O}(\varepsilon^2).
  \label{eq:area_SdS2}
\end{align}
where $\varepsilon = \Lambda B^2$ is also not still considered a small 
dimensionless expansion parameter. Next, expanding the fifth and sixth 
lines in Eq. (\ref{eq:area_SdS2}) with respect to 
$\varepsilon = \Lambda B^2$ and remaining 
${\cal O}((m/B)\cdot \Lambda B^2)$ terms, we have
\begin{align}
 &\quad~
 \iint_{\Sigma^3}\frac{\Lambda}{3}r 
  \left(1 - \frac{\Lambda}{3}r^2\right)^{-\frac{3}{2}}drd\phi
  \nonumber\\
 &=
    - \iint_{\Sigma^3}K^{\rm dS} d\sigma^{\rm dS}
    \nonumber\\
 &+
  \frac{\Lambda B_1 m}{3}\cot\phi_R\csc\phi_R
 -
  \frac{\Lambda B_1 m}{3}\cot\phi_S\csc\phi_S.
  \nonumber\\
 &+
  \frac{\Lambda B_2 m}{3}\cot(\phi_M + \delta_2)\csc(\phi_M + \delta_2)
  -
  \frac{\Lambda B_2 m}{3}\cot(\phi_R + \delta_2)\csc(\phi_R + \delta_2)
  \nonumber\\
 &+
  \frac{\Lambda B_3 m}{3}\cot(\phi_S - \delta_3)\csc(\phi_S - \delta_3)
  -
  \frac{\Lambda B_3 m}{3}\cot(\phi_M - \delta_3)\csc(\phi_M - \delta_3)
  + {\cal O}(\varepsilon^3),
  \label{eq:area_SdS3}
\end{align}
where 
${\cal O}(\varepsilon^3) = {\cal O}
((m/B)^3, (m/B)^2\cdot \Lambda B^2, (m/B)\cdot(\Lambda B^2)^2)$.
Subsequently, we expand the integrand of the second line in Eq. 
(\ref{eq:area_SdS1}) with respect to $\varepsilon = \Lambda B^2$ 
remaining the order ${\cal O}((m/B)^2, (m/B)\cdot\Lambda B^2)$ terms.  
One has
\begin{align}
&\quad~
 \iint_{\Sigma^3}
  \left\{
  \frac{m[6 - \Lambda r^2 (5 - 2\Lambda r^2)]}
  {3r^2 \left(1 - \frac{\Lambda}{3}r^2\right)^{\frac{5}{2}}}
  +
  \frac{m^2[18 - \Lambda r^2 (21 - 10 \Lambda r^2)]}
  {6 r^3 \left(1 - \frac{\Lambda}{3}r^2\right)^{\frac{7}{2}}}
  \right\}drd\phi
  \nonumber\\
  &=
   \iint_{\Sigma^3}\left(\frac{2m}{r^2} + \frac{3m^2}{r^3}\right)drd\phi
   + {\cal O}(\varepsilon^3),
   \label{eq:area_SdS4}
\end{align}
which is of the same form as Eq. (\ref{eq:angle_Sch2}) and
${\cal O}(\varepsilon^3) = 
{\cal O}((m/B)^3, (m/B)^2\cdot \Lambda B^2, (m/B)\cdot(\Lambda B^2)^2)$. 
In this approximation, the order ${\cal O}(m \Lambda)$ 
terms are eliminated from the integrand of Eq. (\ref{eq:area_SdS4}). 
Integrating Eq. (\ref{eq:area_SdS4}) gives 
\begin{align}
 &\quad~
\iint_{\Sigma^3}\left(\frac{2m}{r^2} + \frac{3m^2}{r^3}\right)drd\phi
   \nonumber\\
 &=
  \int_{\phi_R}^{\phi_M}\int_{r^{\rm SdS}_1}^{r^{\rm SdS}_2}
  \left(
   \frac{2m}{r^2} + \frac{3m^2}{r^3} 
			       \right)drd\phi
  + \int_{\phi_M}^{\phi_S}\int_{r^{\rm SdS}_1}^{r^{\rm SdS}_3}
  \left(
   \frac{2m}{r^2} + \frac{3m^2}{r^3} 
  \right)drd\phi
  \nonumber\\
 &=
  2m
  \left[
   \frac{\cos\phi_R - \cos\phi_S}{B_1}
   +
   \frac{\cos (\phi_M + \delta_2) - \cos (\phi_R + \delta_2)}{B_2}
  + 
   \frac{\cos (\phi_S - \delta_3) - \cos (\phi_M - \delta_3)}{B_3}
  \right]
  \nonumber\\
 &- \frac{m^2}{4}
  \left[
   \frac{\sin 2\phi_R - \sin 2\phi_S}{2B_1^2}
   + \frac{\sin 2(\phi_M + \delta_2) - \sin 2(\phi_R + \delta_2)}{2B_2^2}
   \right.\nonumber\\
 &-
    \frac{\sin 2(\phi_M - \delta_3) - \sin 2(\phi_S - \delta_3)}{2B_3^2}
    -\left.
   \frac{\phi_R - \phi_S}{B_1^2}
   + \frac{\phi_M - \phi_R}{B_2^2}
   - \frac{\phi_M - \phi_S}{B_3^2}
   \right)
   + {\cal O}(\varepsilon^3),
   \label{eq:area_SdS5}
\end{align}
which corresponds to the Schwarzschild-like part given in 
Eqs. (\ref{eq:alpha_Sch1}) and (\ref{eq:angle_Sch2}).

Substituting Eqs. (\ref{eq:area_SdS2}), (\ref{eq:area_SdS3}), 
and (\ref{eq:area_SdS5}) into Eq. (\ref{eq:area_SdS1}) and
applying to Eq. (\ref{eq:deftotalangle4}), we have
\begin{align}
 \alpha_{\rm SdS} 
  &= \left|
       \iint_{\Sigma^3}K^{\rm SdS} d\sigma^{\rm SdS}
       - 
       \iint_{\Sigma^3}K^{\rm dS} d\sigma^{\rm dS}
      \right|
  \nonumber\\
 &=
  2m
  \left[
   \frac{\cos\phi_R - \cos\phi_S}{B_1}
   +
   \frac{\cos (\phi_M + \delta_2) - \cos (\phi_R + \delta_2)}{B_2}
   +
   \frac{\cos (\phi_S - \delta_3) - \cos (\phi_M - \delta_3)}{B_3}
  \right]
  \nonumber\\
 &- \frac{m^2}{4}
  \left[
   \frac{\sin 2\phi_R - \sin 2\phi_S}{2B_1^2}
   + \frac{\sin 2(\phi_M + \delta_2) - \sin 2(\phi_R + \delta_2)}{2B_2^2}
   \right.\nonumber\\
 &-
   \frac{\sin 2(\phi_M - \delta_3) - \sin 2(\phi_S - \delta_3)}{2B_3^2}
   -\left.
   15\left(
   \frac{\phi_R - \phi_S}{B_1^2}
   + \frac{\phi_M - \phi_R}{B_2^2}
   - \frac{\phi_M - \phi_S}{B_3^2}
   \right)
    \right]\nonumber\\
 &+
  \frac{\Lambda B_1 m}{3}\cot\phi_R\csc\phi_R
 -
  \frac{\Lambda B_1 m}{3}\cot\phi_S\csc\phi_S.
  \nonumber\\
 &+
  \frac{\Lambda B_2 m}{3}\cot(\phi_M + \delta_2)\csc(\phi_M + \delta_2)
 -
  \frac{\Lambda B_2 m}{3}\cot(\phi_R + \delta_2)\csc(\phi_R + \delta_2)
  \nonumber\\
 &+
  \frac{\Lambda B_3 m}{3}\cot(\phi_S - \delta_3)\csc(\phi_S - \delta_3)
 -
  \frac{\Lambda B_3 m}{3}\cot(\phi_M - \delta_3)\csc(\phi_M - \delta_3)
 \nonumber\\
  &+
  {\cal O}(\varepsilon^3).
  \label{eq:defangleSdS2}
\end{align}
Eq. (\ref{eq:defangleSdS2}) completely agrees with 
Eq. (\ref{eq:defangleSdS1}).

Before closing this section, it is noteworthy that in the case of 
asymptotically flat spacetime, we can say that there is no ambiguity 
about what is meant by the total deflection angle when the source and 
observer are sufficiently far away. However in the case of non-asymptotically 
flat spacetime, the total deflection angle depends on the choice of 
the triangle because different triangles in curved spacetime cause different 
sums of interior angles. Therefore, it is important to clarify how 
the triangle is constructed in a non-asymptotically flat spacetime.

\subsection{Contribution of the Cosmological Constant and its Observability}
Let us investigate how the cosmological
constant $\Lambda$ contributes to the total deflection angle, and
its observability.

Using Eqs. (\ref{eq:defangleSdS1}) and (\ref{eq:defangleSdS2}),
we extract the part of the order ${\cal O}(\Lambda m)$ terms 
and set
\begin{align}
 \alpha_{\rm SdS}^{\Lambda}
  &=
  \frac{\Lambda B_1 m}{3}\cot\phi_R\csc\phi_R
  -
  \frac{\Lambda B_1 m}{3}\cot\phi_S\csc\phi_S.
  \nonumber\\
 &+
  \frac{\Lambda B_2 m}{3}\cot(\phi_M + \delta_2)\csc(\phi_M + \delta_2)
 -
  \frac{\Lambda B_2 m}{3}\cot(\phi_R + \delta_2)\csc(\phi_R + \delta_2)
  \nonumber\\
 &+
  \frac{\Lambda B_3 m}{3}\cot(\phi_S - \delta_3)\csc(\phi_S - \delta_3)
 -
  \frac{\Lambda B_3 m}{3}\cot(\phi_M - \delta_3)\csc(\phi_M - \delta_3).
  \label{eq:defangleSdS3}
\end{align}
We observe that the cosmological constant $\Lambda$ contributes to
the total deflection angle, and that the leading terms have a form of 
${\cal O}(\Lambda m)$ instead of ${\cal O}(\Lambda/m)$.
Furthermore, the terms characterized only by the cosmological constant 
$\Lambda$ do not appear in the expression of the total deflection 
angle $\alpha_{\rm SdS}$ 
(see Eqs. (\ref{eq:angle_dS1}) and (\ref{eq:angle_dS2})). 
This is because angle $\psi^{\rm SdS}$ and the areal integral of 
Gaussian curvature $K^{\rm SdS}$ can be expressed as Eqs. 
(\ref{eq:SdScorrectionterms1}) and (\ref{eq:SdScorrectionterms2}); 
therefore, the terms are completely eliminated, as seen in Eqs. 
(\ref{eq:defangleSdS1}) and (\ref{eq:defangleSdS2}).

Let us discuss the observability of the cosmological
constant $\Lambda$ to the total deflection angle.  
We assume $\Lambda \approx 10^{-52} ~ {\rm m}^{-2}$
and consider the sun $m_{\odot} = GM_{\odot}/c^2$ and typical galaxy 
$m_{\rm gal} \approx 10^{12}GM_{\odot}/c^2$ as the lens objects,
where $G = 6.674 \times 10^{-11} ~{\rm m^3 \cdot kg^{-1} \cdot s^{-2}}$
denotes the Newtonian gravitational constant, 
$c = 3.0 \times 10^{8}~{\rm m}^2$ 
the speed of light in vacuum, and 
$M_{\odot} = 2.0 \times 10^{30} ~{\rm kg}$ the mass of the sun.
Additionally, we employ the radius of sun and galaxy as the impact 
parameters 
$b_1 \approx b_{\odot} \approx B_{\odot} \approx R_{\odot} 
= 6.960 \times 10^8~{\rm m}$ 
and $b_1 \approx b_{\rm gal} \approx B_{\rm gal} \approx R_{\rm galaxy} 
\approx 5.0 \times 10^4 ~ {\rm ly} \approx 5 \times 10^{20}~{\rm m}$, 
respectively. 

When the lens object is the sun, we observe
\begin{align}
 \frac{4m_{\odot}}{B_{\odot}} \approx 8.5 \times 10^{-6},\quad
  \frac{15\pi m^2_{\odot}}{4B_{\odot}^2} \approx 5.3 \times 10^{-11},\quad
  \frac{\Lambda B_{\odot}m_{\odot}}{3} \approx 3.3 \times 10^{-41},
\end{align}
where the unit is in rad (radian). The order of the contribution of 
the cosmological constant $\alpha_{\rm SdS}^{\Lambda}$ is 
${\cal O}(10^{-41}) ~~ {\rm rad}$. Even if we consider 
the $\cot\phi\csc\phi$ term, $\alpha_{\rm SdS}^{\Lambda}$ is only 
${\cal O}(10^{-36})$ at most where we assumed that observer 
$R$ and source $S$ are located at the position of Earth and its opposition,
i.e., $\phi_R = \arcsin B_{\odot}/(1.5 \times 10^{11})$, and
$\phi_S = \pi - \arcsin B_{\odot}/(1.5 \times 10^{11})$, respectively.

However, when the lens object is the galaxy, we observe
\begin{align}
 \frac{4m_{\rm gal}}{B_{\rm gal}} \approx 1.2 \times 10^{-5},\quad
  \frac{15\pi m^2_{\rm gal}}{4B_{\rm gal}^2} \approx 1.0 \times 10^{-10},\quad
  \frac{\Lambda B_{\rm gal}m_{\rm gal}}{3} \approx 2.5 \times 10^{-17}.
\end{align}
Accordingly, $\alpha_{\rm SdS}^{\Lambda}$ is 
$\alpha_{\rm SdS}^{\Lambda} \approx {\cal O}(10^{-17})$ and
mostly three orders of magnitude smaller than the sensitivity of 
the angle observed in planned space missions (such as LATOR), 
$0.01~{\rm picorad} = 10^{-14} ~{\rm rad}$
(see Figure 3 in \cite{lator2009}). 

However, because of the $\cot\phi\csc\phi$ term, 
$\alpha_{\rm SdS}^{\Lambda}$ 
rapidly increases when source $S$ and receiver $R$ reach
the de Sitter horizon $r \rightarrow r_{\Lambda} = \sqrt{3/\Lambda}
\approx 1.73 \times 10^{26} ~{\rm m}$, where we estimated the angular 
coordinate of the de Sitter horizon as
$\phi_{\Lambda} = \arcsin(B_{\rm gal}/r_{\Lambda}) \approx 3.0 \times 10^{-6}$,
and suppose triangle $\Sigma^3$ to be symmetrical with respect 
to $\phi = \pi/2$ (see Figures \ref{fig:arakida-fig4} and 
\ref{fig:arakida-fig5}). Let source $S$ and receiver $R$ located in
a symmetrical position near the de Sitter horizon with 
$\phi_S = \phi_{\Lambda}$ and $\phi_R = \pi - \phi_{\Lambda}$. 
Furthermore, because of the same reason as that in Eq. (\ref{eq:Schlimit2}),
$S$ and $R$ approach the de Sitter horizon, and $b_2 = b_3$ also approaches 
the de Sitter horizon,  i.e., $b_2 = b_3 \rightarrow r_{\Lambda}$. 

However, it is reasonable to set  
\begin{align}
 \cot \left(\frac{\pi}{2} - \delta_2\right)
 \csc \left(\frac{\pi}{2} - \delta_2\right) 
 &= -
 \cot \left(\frac{\pi}{2} + \delta_2\right)
 \csc \left(\frac{\pi}{2} + \delta_2\right) 
 \simeq {\cal O}(1)\\ 
 \cot (\phi_{\Lambda} + \delta_2)\csc (\phi_{\Lambda} + \delta_2)
 &\simeq \cot\delta_1\csc\delta_2
 \simeq {\cal O}(10) \sim {\cal O}(10^2)\\
 \cot (\pi - \phi_{\Lambda} - \delta_2)
 \csc (\pi - \phi_{\Lambda} - \delta_2) 
 &\simeq
 \cot (\pi - \delta_2)\csc (\pi - \delta_2)
 \simeq {\cal O}(10) \sim {\cal O}(10^2),
\end{align}
where we assumed $5 \lesssim \delta_2 \lesssim 30$ degree.
Then, we can estimate as
\begin{align}
 &
 \frac{\Lambda b_2 m_{\rm gal}}{3}
 \cot \left(\frac{\pi}{2} + \delta_2\right)
 \csc \left(\frac{\pi}{2} + \delta_2\right) 
 = -
 \frac{\Lambda b_2 m_{\rm gal}}{3}
 \cot \left(\frac{\pi}{2} - \delta_2\right)
 \csc \left(\frac{\pi}{2} - \delta_2\right)
 \simeq 10^{-11},\\
 &
\frac{\Lambda b_2 m_{\rm gal}}{3}
 \cot \delta_2 \csc \delta_2
 =
 -
 \frac{\Lambda b_2 m_{\rm gal}}{3}
 \cot (\pi - \delta_2)\csc (\pi - \delta_2)
 \simeq 10^{-10} \sim 10^{-9}.
\end{align}
Accordingly, $\alpha^{\Lambda}_{\rm SdS}$ can be evaluated as
\begin{align}
 \alpha^{\Lambda}_{\rm SdS} \approx - 5.9 \times 10^{-6}
  ~{\rm rad},
\end{align}
which is almost half the value of the Schwarzschild part.
Hence, if both source $S$ and receiver $R$ are located near the
de Sitter horizon, we may be able to detect the contribution of
the cosmological constant $\Lambda$ to the total deflection angle. 

\section{Summary and Conclusions \label{sec:conclusion}}
Assuming a static and spherically symmetric spacetime,
we proposed a new concept of the total deflection angle 
of a light ray in curved spacetime by means of the 
optical geometry which is considered as the Riemannian geometry 
experienced by the light ray. The concept is defined by 
the difference between the sum of internal angles of two triangles; 
one of them lies on curved spacetime distorted by a 
gravitating body and the other on the background spacetime.
Our new definition of the total deflection angle is geometrically 
and intuitively clear. The triangle used to define 
the total deflection angle was realized by setting three laser-beam 
baselines (i.e., three null geodesics) in the space, inspired 
from planned space missions, including, LATOR, ASTROD-GW, and LISA. 
Accordingly, our new total deflection angle can be calculated by 
measuring the difference between the sum of the internal angles 
of both the triangles.
Two formulas were presented to calculate the total deflection 
angle, in accordance with the Gauss--Bonnet theorem.
It was shown that in the case of the Schwarzschild spacetime,
the expression of the total deflection angle reduced to 
Epstein--Shapiro's formula when the source of the light ray, 
$S$, and observer $R$ were located in an asymptotically flat region. 
Additionally, in the case of the Schwarzschild--de Sitter spacetime, 
the total deflection angle was represented by the Schwarzschild-like 
part and the coupling terms of the central mass $m$ and cosmological 
constant $\Lambda$ as the form of ${\cal O}(\Lambda m)$ instead of 
${\cal O}(\Lambda/m)$. 
Furthermore, the expression for the total deflection angle 
did not include the terms characterized solely by the 
cosmological constant $\Lambda$; 
this is obvious from the fact that the angle 
$\psi^{\rm SdS}_p$ and the area integrals of the Gaussian curvature 
$K^{\rm SdS}$ in the Schwarzschild--de 
Sitter spacetime can be expressed using those in the de Sitter spacetime, 
$\psi^{\rm dS}_p$ and $K^{\rm dS}$ as described in Eqs 
(\ref{eq:psi_1_SdS}), (\ref{eq:psi_2_SdS}), (\ref{eq:psi_3_SdS}),
and (\ref{eq:area_SdS3}), see also 
(\ref{eq:SdScorrectionterms1}) and (\ref{eq:SdScorrectionterms2}).

When the lens object was the sun, the magnitude of the
contribution of the cosmological constant to the total deflection
angle was $\alpha_{\rm SdS}^{\Lambda} \simeq {\cal O}(10^{-36})$. 
Accordingly, it was extremely difficult to detect the contribution. 
However, if the galaxy was the lens object, the contribution of the 
cosmological constant was approximately
$\alpha_{\rm SdS}^{\Lambda} \simeq {\cal O}(10^{-17})$, which is 
mostly three orders of magnitude smaller than the sensitivity 
$0.01 ~ {\rm picorad} = 10^{-14} ~ {\rm rad}$ observed in 
planned space missions such as LATOR. However, when the source of 
the light ray, $S$, and the observer $R$ reached the de Sitter 
horizon, $\alpha_{\rm SdS}^{\Lambda}$ became 
$\alpha_{\rm SdS}^{\Lambda} \simeq -5.9 \times 10^{-6} ~ {\rm rad}$,
which is almost half of the Schwarzschild part.
Therefore, if we can observe gravitational lensing of 
distant galaxies such as the recently reported RXCJ0600-z6
\cite{fujimoto_etal2021} which is located almost at the de Sitter horizon,
$12.9 ~{\rm billion\ ly} \simeq 1.22 \times 10^{26} ~{\rm m} ~ (z = 6.0719)$,
it may be possible to detect the contribution of the cosmological constant 
to the bending of light using the gravitational lensing effect.

Especially for observations in the solar system, the clocks on board 
the three satellites need to be synchronized accurately. Then the problem 
of synchronization of the clocks must be discussed as a future problem 
when our results are applied to actual measurements.

Gravitational lensing may pave the way to solving the cosmological 
constant problem. However, it is currently difficult 
to use our formulation directly for cosmological constant/dark energy 
exploration using gravitational lensing, and further extensions and 
improvements are required. This is because first we must arrange 
the polygons such that the singularity of the central object 
is successfully incorporated into the gravitational lensing equation.
And second, in our formulation we need to place the observer at 
the three vertices of the triangle. However, in the actual gravitational 
lensing effect, the observer can only be placed at one of the three vertices 
of the triangle
\footnote{
Nevertheless, although there are some problems in practical applications, 
we think that our results are useful because we can show 
how the formula of the total deflection angle of light ray is 
affected by the cosmological constant without ambiguity.
}.
Therefore, we have also to resolve these problems in the future.


\acknowledgments

We would like to acknowledge anonymous referee for reading our paper 
carefully and for giving fruitful comments and suggestions, 
which significantly improved the quality of the paper.
We also appreciate H. Asada and K. Takizawa for fruitful discussions 
and comments. This work was partially supported by Dean's Grant for 
Specified Incentive Research, College of Engineering, Nihon University.


\end{document}